\documentclass[useAMS,usenatbib]{aa}  

\usepackage{graphicx}
\usepackage{txfonts}
\usepackage{hyperref}
\usepackage{longtable}
\usepackage{appendix}
\begin{document}

   \title{A second low-mass planet orbiting the nearby M-dwarf GJ 536}

   \author{A.~Su{\'a}rez~Mascare{\~n}o\inst{\ref{iac},\ref{uiac}},
          C.~del~Burgo\inst{\ref{iac},\ref{uiac}},
          J.-B.~Delisle\inst{\ref{ugeneva}},
          J.~I.~Gonz\'alez~Hern\'andez\inst{\ref{iac},\ref{uiac}},
          N.~C.~Hara\inst{\ref{umar}},
          J.~M.~Mestre\inst{\ref{padova}},
          N.~Nari\inst{\ref{lb},\ref{iac},\ref{uiac}},          
          R.~Rebolo\inst{\ref{iac},\ref{uiac},\ref{csic}},
          A.~K.~Stefanov\inst{\ref{iac},\ref{uiac}}, 
          and J.~A.~Burt\inst{\ref{JPL}}
          }

    \authorrunning{A.~Su{\'a}rez Mascare{\~n}o et al.}
    
   \institute{Instituto de Astrof\'{\i}sica de Canarias, c/ V\'ia L\'actea s/n, 38205 La Laguna, Tenerife, Spain\label{iac}\\
    \email{asm@iac.es} 
    \and Departamento de Astrof\'{\i}sica, Universidad de La Laguna, 38206 La Laguna, Tenerife, Spain \label{uiac}
    \and Universit\'{e} Aix Marseille, CNRS, CNES, LAM, Marseille, France \label{umar}
    \and Observatoire Astronomique de l'Universit\'{e} de Gen\`{e}ve, 51 Chemin de Pegasi, 1290 Versoix, Switzerland  \label{ugeneva}
    \and Light Bridges S.L., Observatorio del Teide, Carretera del Observatorio, s/n Guimar, 38500, Tenerife, Canarias, Spain \label{lb}
    \and Università di Padova, Vicolo dell'Osservatorio 3, I-35122 Padova, Italy \label{padova}
    \and Consejo Superior de Investigaciones Cient\'ificas (CSIC), E-28006 Madrid, Spain \label{csic}
    \and Jet Propulsion Laboratory, California Institute of Technology, 4800 Oak Grove Drive, Pasadena, CA 91109, USA \label{JPL}
    }

   \date{Written March-May 2025}

 
  \abstract
   {GJ 536 is a low-mass star, located 10 pc away from the Sun, that hosts a low-mass planet orbiting with a period of 8.71 days. Based on an analysis of the radial velocity (RV) time series obtained from the available data of the spectrographs HARPS, \mbox{HARPS-N}, CARMENES and HIRES, we announce the discovery of a second low-mass planet orbiting the star. We performed a RV global analysis on RV, spectroscopic activity indicators, and ASAS photometry, within the multidimensional Gaussian process framework, updated the parameters of GJ 536 b, and found significant evidence of the presence of a second planet. GJ 536 c is a low-mass planet ($m_{p} \sin i$ = 5.89 $\pm$ 0.70 M$_{\oplus}$), orbiting with a period of 32.761 $\pm$ 0.015 days, at a distance of 0.1617 $\pm$ 0.0028 au from its parent star. It induces an RV semi-amplitude of 1.80 $\pm$ 0.20 m$\cdot$s$^{-1}$. Given its distance to the star, it receives a flux of 1.692 $\pm$ 0.069 F$_{\oplus}$, for an equilibrium temperature of 290.5 $\pm$ 9.5 K. We update the mass of the planet GJ 536 b to $m_{p} \sin i$ = 6.37 $\pm$ 0.38 M$_{\oplus}$. The orbits of both planets are consistent with circular. We explored the use of statistical Doppler imaging on the photometric and RV data, and find a tentative projected obliquity of the stellar rotation axis of 58$^{+16}_{-19}$ deg. Current evidence does not support the presence of additional planets with masses $\textgreater$ 5 M$_{\oplus}$for orbital periods up to 100 days, or $\textgreater$ 10 M$_{\oplus}$ for periods up to 1000 days.}

   \keywords{exoplanets --
                radial velocity --
                transits -- 
                stellar activity --
                super-Earths
               }

   \maketitle

\section{Introduction}

The detection of low-mass exoplanets has been fueled by technological and methodological advancements. The arrival of high-precision, ultra-stabilised spectrographs, opened the possibility of detecting low-mass exoplanets \citep{Santos2004, Mayor2009}. Later, improvements in RV determination yielded better RV precision than what was originally expected to be achievable \citep{AngladaEscude2016}. Stellar activity is nowadays the main limiting factor when attempting to detect planets that induce m$\cdot$s$^{-1}$ RV signals \citep{Robertson2014, Masca2017b}. In recent years, the techniques used to correct stellar activity have progressed significantly, enabling the detection of planets inducing signals smaller than those of stellar activity \citep{Haywood2014,Faria2021}. Simultaneously, new statistical methods have been developed to optimally incorporate the stellar variability model in the computation of the statistical significance of candidate planets \citep{hara2022a}. With all these tools at hand, it is now easier to identify the presence of very low-mass planets, and asses their planetary nature. A consequence of these improvements in radial velocity extraction and modelling tools is that it is now possible to find planets hidden in archival data. 

GJ 536 is a nearby M1V dwarf that hosts a close-in low-mass planet \citep{Masca2017}. The planet was identified using a traditional periodogram analysis on the HARPS and HARPS-N RV data, derived from cross-correlation functions (CCFs). The rotation period of the star was measured as 43 days, and modelled using a double sinusoidal model. At the time, no other signals were identified in the data. 

In this paper, we report the detection of a second low-mass planet orbiting inner to the edge of the optimistic habitable zone of the star. We perform a re-extraction of the RVs using a Line-By-Line algorithm, we complete the RV data using the publicly available CARMENES and HIRES data, perform a global model using multidimensional Gaussian processes (GP) regression, and asses the detections using state-of-the-art statistical models. 

\section{Observations and data} \label{obs_data}

This work is based on archival observations from the High Accuracy Radial velocity Planet Searcher \citep[HARPS,][]{Mayor2003}, HARPS-N \cite{Cosentino2012}, the Calar Alto high-Resolution search for M-dwarfs with Exoearths with Near-infrared and optical Échelle Spectrographs \citep[CARMENES,][]{Quirrenbach2014}, and The High Resolution Echelle Spectrometer \citep[HIRES,][]{Vogt1994}.

HARPS and HARPS-N are fibre-fed high-resolution echelle spectrographs with similar resolving power and spectral range, $R\sim 115\,000$ between $\sim$380 and $\sim$690 nm and have been designed to attain very high long-term radial-velocity precision. HARPS is installed at the 3.6-m ESO telescope in La Silla Observatory (Atacama, Chile), while HARPS-N is installed at the 3.6-m Telescopio Nazionale Galileo of the Roque de los Muchachos Observatory (La Palma, Spain). Both are contained in temperature- and pressure-controlled vacuum vessels to avoid spectral drifts due to temperature and air pressure variations, thus  ensuring its stability. They are equipped with their own pipeline providing extracted and wavelength-calibrated spectra, as well as RV measurements and other data products such as cross-correlation functions and their bisector profiles.

CARMENES consists of visual (VIS) and near-infrared (NIR) vacuum-stabilised spectrographs covering 520 -- 960 nm and 960 -- 1710 nm with a spectral resolution of 94~600 and 80~400, respectively. It is located at the 3.5-m Zeiss telescope at the Centro Astronómico Hispano Alemán (Almería, Spain).  The spectra are extracted with the \texttt{CARACAL} pipeline, based on flat-relative optimal extraction \citep{Zechmeister2014}. The wavelength calibration was performed by combining hollow cathode (U-Ar, U-Ne, and Th-Ne) and Fabry-P\'erot etalons exposures. The instrument drift during the nights is tracked with the Fabry-P\'erot in the simultaneous calibration fibre. 

HIRES is a slit-fed, grating cross-dispersed echelle spectrograph with a spectral resolution of 55 - 80~000 depending on the input slit decker \citep{Vogt1994}. HIRES is installed at the Keck I telescope on Mauna Kea (Hawaii, USA) and uses an iodine cell placed in the converging beam of the telescope as a wavelength calibrator \citep{MarcyButler1992}. This cell imprints incoming stellar light with a dense forest of I2 lines from 500 - 620 nm, providing a precise wavelength solution in this region. The I2 lines also act as a proxy for the point spread function (PSF) of the instrument, and so changes to the instrument profile caused by temperature or pressure variations will be reflected in the observed absorption line profiles \citep{Butler1996}. While HIRES has a spectral range of 370-800 nm, only this 120 nm iodine region is used when measuring precise radial velocities.

In addition to the spectroscopic data, we used the photometric time series obtained by the All Sky Automated Survey (ASAS; \citealt{Pojmanski1997}). The ASAS data is composed of $V$-band observations taken by the ASAS-3 survey, obtained from the ASAS website~\footnote{https://www.astrouw.edu.pl/asas/}. The light curves include the photometric measurements obtained using four different apertures. We averaged over the four apertures to obtain the final photometric series, reducing the short-term scatter of the data. 

Last, GJ 536 has been observed by the All Sky Automated Survey for SuperNovae (ASAS-SN; \citealt{Kochanek2017}) and the Transiting Exoplanet Survey Satellite (\textit{TESS}) \citep{Ricker2015}.The default aperture photometry data of ASAS-SN were strongly contaminated by the moon. We attempted to clean them by using the saturated stars photometry method \citep{Winecki2024} and removing epochs with lunar separations of less than 90 degrees. The resulting time series showed no evidence of astrophysical variations. \textit{TESS} observed GJ~536 in sector 50. We found no evidence of transits in the \textit{TESS} data (see Appendix~\ref{tess_phot}). 

\subsection{Radial velocities}

We extracted the HARPS and HARPS-N RVs using the Line-By-Line (\texttt{LBL}\footnote{version 0.65.001, lbl.exoplanets.ca}) code developed by \citet{Artigau2022}, and based on \citet{Dumusque2018}. The \texttt{LBL} algorithm performs an outlier-resistant template matching to each individual line in the spectra. For non-telluric corrected spectra, it produces its own telluric correction. In addition to the velocity, it derives other quantities such as a differential line-width (dLW, similar to \citealt{Zechmeister2018}). The dLW is obtained from the second derivative of the template and can be understood as a change in the line full width at half maximum (FWHM), assuming a Gaussian profile. Using the dLW variations, and an estimation of the average FWHM of the lines, the \texttt{LBL} algorithm estimates a variation of the FWHM of the lines. We used these measured changes in the FWHM as our main activity indicator. In addition, the \texttt{LBL} algorithm derives a chromatic RV slope over the full spectral range (chromatic index, CRX,  similar to \citealt{Zechmeister2018}). In addition, we used the Barycentric Earth Radial Velocity (BERV) to track potential leftover effects from the telluric correction.

The time series data of CARMENES were taken from the CARMENES DR1 \citep{Ribas2023}. These RVs were obtained using \texttt{SERVAL}. This software builds a high signal-to-noise template by co-adding all the existing observations, and then performs a maximum likelihood fit of each observed spectrum against the template, yielding a measure of the Doppler shifts and their uncertainty. The time series provided includes additional indicators, such as the FWHM of a CCF built using the \texttt{RACOON} pipeline \citep{Lafarga2020}.

We used the HIRES RV time series provided by \citet{Rosenthal2021}, which, in addition, included time series of Mount Wilson $S$-index. The HIRES RVs were obtained by measuring the Doppler shift of starlight relative to a reference spectrum of molecular iodine, which is at rest in the observatory frame \citep{Butler1996}. 

\subsection{Activity proxies}

The presence of active regions on the stellar surface affects the flux emitted by the star and its velocity field, distorting the shape of the lines and the point spread function measured by the instrument \citep{Dravins1982}. To measure these variations, we use the full width at half maximum (FWHM), provided by the \texttt{LBL} code and provided by the CARMENES DR1. 

The intensity of the emission of the cores of the Ca II H\&K lines is linked to the strength of the magnetic field of the star, which in turn is well correlated with the stellar rotation period for FGKM stars \citep{Noyes1984}. The measured emission intensity of the line cores also changes when active regions move across the stellar disc, helping us trace the rotation of the star. We calculate the Mount Wilson $S$ index ($S_{\rm MW}$) for the HARPS and HARPS-N data following \citet{Vaughan1978}. We define two triangular-shaped passbands with a FWHM of 1.09~{\AA} centred at 3968.470~{\AA} and at 3933.664~{\AA} for the Ca II H\&K line cores. For the continuum, we use two bands 20~{\AA} in width centred at 3901.070~{\AA} (V) and 4001.070~{\AA} (R).
Then the Snindex is defined as: 

\begin{equation}
   S=\alpha {{\tilde{N}_{H}+\tilde{N}_{K}}\over{\tilde{N}_{R}+\tilde{N}_{V}}} + \beta,
\end{equation}
\noindent where $\tilde{N}_{H},\tilde{N}_{K},\tilde{N}_{R}$, and $\tilde{N}_{V}$ are the mean fluxes per wavelength unit in each passband,  while $\alpha$ and $\beta$ are calibration constants fixed at $\alpha = 1.111$ and $\beta = 0.0153$ \citep{Lovis2011}. The $S$ index serves as a measurement of the Ca II H\&K core flux normalised to the neighbour continuum. 

The HIRES S$_{MW}$ measurements from \citet{Rosenthal2021} are provided without estimates for their uncertainties. We downloaded the original spectra and computed the photon noise-uncertainties by propagating the photon noise in each band, as with the HARPS and HARPS-N data (e.g. \citealt{Laliotis2023}).

We complement the data with the photometric time series coming from the All Sky Automated Survey (ASAS; \citealt{Pojmanski1997}). We used the observations obtained by the ASAS-3. It consists of $V$-band data taken over a long baseline. We obtained the ASAS-3 data from the ASAS website~\footnote{https://www.astrouw.edu.pl/asas/}. The light curves include the photometric measurements obtained using four different apertures. We averaged over the four apertures to obtain the final photometric series, which reduced the short-term scatter of the data.

\subsection{Summary of time series}

\begin{figure}[!ht]
   \centering
  \includegraphics[width=9cm]{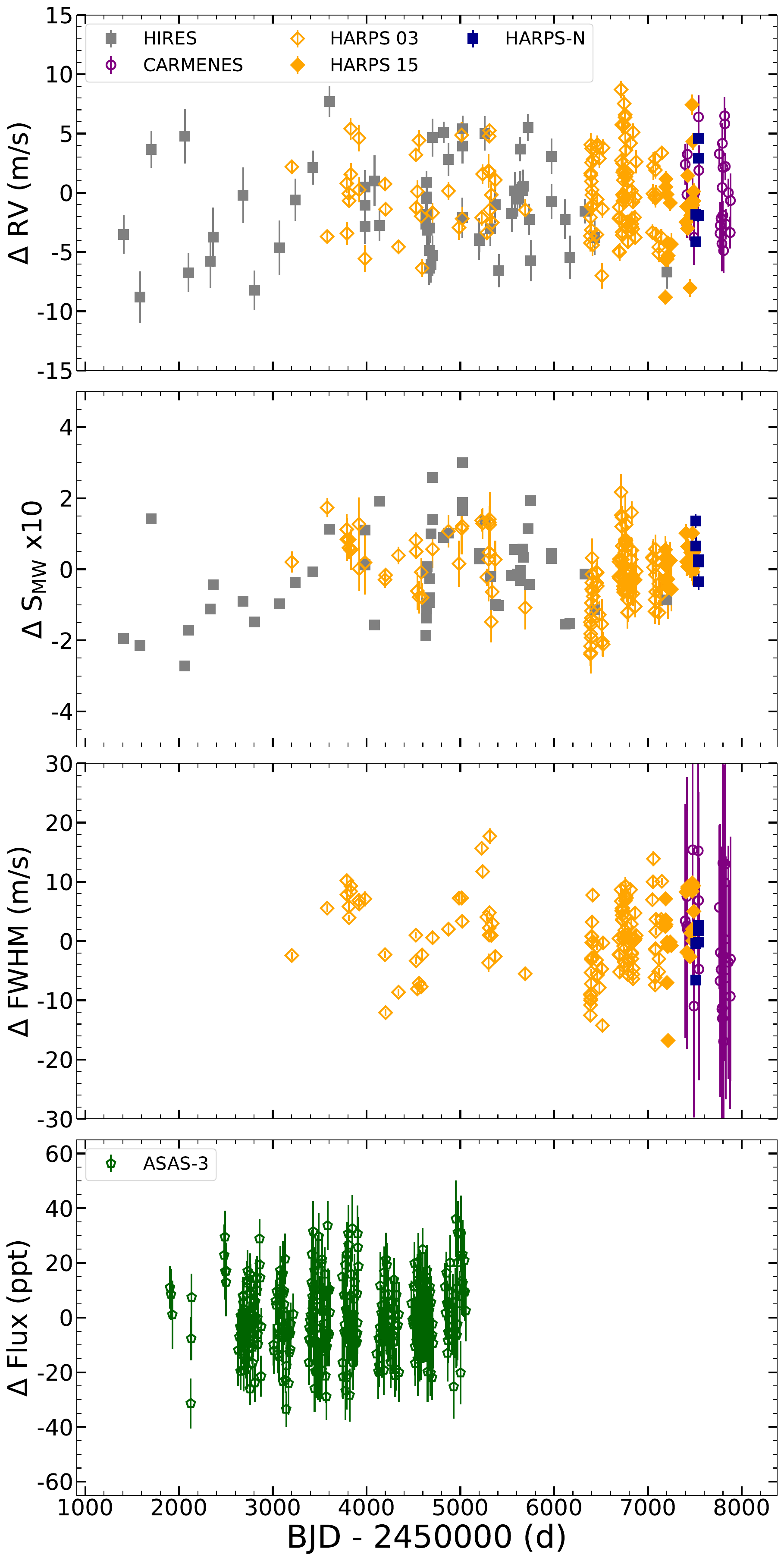}
  \caption{Spectroscopic and photometric data. Time series of RV, S$_{MW}$, FWHM, and Flux$_{V}$ used in this study.}
  \label{data}
\end{figure}

\begin{table*}[!ht]
   \begin{center}
       \caption{RV data used in this work \label{tab:data}}
       \begin{tabular}[center]{l l l l l l}
           \hline
            & HARPS & HARPS-N & CARMENES & HIRES & Combined \\
           \hline    
           Extraction & \texttt{LBL} & \texttt{LBL} &  \texttt{SERVAL} & Iodine cell \\
           N. data & 147 & 5 & 25 & 58 & 235 \\
           Baseline [days] & 5286 & 30 & 465 & 5793 & 6469 \\     
           RV RMS [m$\cdot$s$^{-1}$] & 3.34 & 3.28 & 3.4 & 3.9 & 3.55 \\
           $\sigma$RV [m$\cdot$s$^{-1}$]& 0.69 & 0.37 & 1.4 & 1.5 & 0.81 \\ 
           FWHM RMS [m$\cdot$s$^{-1}$] & 6.22 & 3.24 & 9 &  & 6.68 \\
           $\sigma$FWHM [m$\cdot$s$^{-1}$]& 0.75 & 0.36 & 20 &  & 0.81 \\ 
           S$_{\rm MW}$ RMS [dex] & 0.088 & 0.056 &  & 0.123 & 0.098 \\
           $\sigma$S$_{\rm MW}$  [dex]& 0.038 & 0.024 &  & 0.038$^{*}$ & 0.038 \\ 
           \hline
       \end{tabular}
   \end{center}
   $^{*}$Assigned value
\end{table*}

Overall, we analysed 235 nightly binned RV measurements, distributed as 147 from HARPS, 5 from HARPS-N, 25 from CARMENES, and 58 from HIRES. The combined RV time series shows an RMS of 3.55 m$\cdot$s$^{-1}$, and a median uncertainty of 81 cm$\cdot$s$^{-1}$. We analysed 177 FWHM measurements from HARPS, HARPS-N, and CARMENES. The combined FWHM time series shows an RMS of 6.68 m$\cdot$s$^{-1}$ with a median uncertainty of 81 cm$\cdot$s$^{-1}$. We analysed 210 S$_{\rm MW}$ measurements from HARPS, HARPS-N, and HIRES. The data shows an RMS of 0.983 with a median uncertainty of 0.038. Last, we analysed 325 photometric epochs, with an RMS of 13.1 parts-per-thousand (ppt) and a median uncertainty of 8.3 ppt. Figure~\ref{data} shows the time series of RV, S$_{\rm MW}$ index, FWHM, and V-fluxes. Figure~\ref{data} shows the time series of RV, S$_{\rm MW}$ index, FWHM, and V-fluxes. Table~\ref{tab:data} shows a summary of the spectroscopic data considered for the analysis, with details for every individual instrument.  We found the RV data to show weak levels of corelationship with the S$_{\rm MW}$ and FWHM and data. More details can be found in Appendix~\ref{append_correl}.

\section{Fundamental stellar parameters of GJ 536}

We derived the fundamental stellar parameters of GJ\ 536 using the Bayesian inference code of \citet{delburgo2016,delburgo2018} on the PARSEC v1.2S library of stellar evolution models \citep{bressan2012, Chen2014, Tang2014, Chen2015}, which shows good statistical agreement with the dynamical masses of main-sequence detached eclipsing binaries (within \,4~\%, \citealt{delburgo2018}). 

We arranged a grid of stellar evolution models with ages ranging from 2 to 13~800 million years and steps of 5\%, and [M/H] from --2.18 to 0.51 with steps of 0.02 dex, adopting the photometric passband calibration of \citet{Riello2021}, with the zero points of the VEGAMAG system. We then applied the Bayesian method described in \citet{delburgo2016}, taking as imputs the absoltute G magnitude and the colour $G_{BP}-G_{RP}$, the absolute $G$ magnitude $M_{G}$  from \textit{Gaia} DR3 \citep{GaiaEDR3}, assuming a null extinction, and the abundance ratio of iron to hydrogen [Fe/H]=$-0.08~\pm~0.09$ \citep{Maldonado2020}. With this method, we inferred the stellar mass, radius, luminosity, effective temperature, and surface gravity. It is important to note that the uncertainties of derived stellar masses and radii are typically smaller than the dispersion between different models. As these parameters can affect the computation of planetary parameters, we enlarged the error bars of the stellar parameters in all our calculations following \citet{Tayar2022}. The final uncertainty in the stellar mass of \,4~\% is consistent with the measured statistical accuracy of the code. Table \ref{tab:stellar_parameters} shows the inferred parameters of GJ\ 536. 

We measured a rotational period of 43.63 $\pm$ 0.85 d, which can be used for dating GJ\ 536 from the gyrochronology-based relation of \citet{Mamajek2008}, taking into account the colour B-V= 1.47 of this red dwarf \citep{Koen2010}. The resulting gyro-kinematic age is 4.2 $\pm$ 0.7 Ga, which is lower but nearly within 1-$\sigma$ when compared to the isochronal age, which matches that of \citet{Maldonado2020} (8$\pm$4 Ga). 

GJ 536 has a magnetic cycle originally reported by \citet{Masca2017}, with a period of $\sim$ 825 days, and recently updated by \citet{IbanezBustos2025} to a period of $\sim$ 1600 days, with a second component at longer period ($\sim$ 4000 days).

Using the reported stellar parameters, we estimate the limits of the habitable zone (HZ) of the star to be 0.176 -- 0.372 au, following \citet{Kopparapu2014, Kopparapu2017}, for the runaway greenhouse to early-Mars limits. These limits correspond to orbital periods of 37.1 $\pm$ 1.1 days to 114.2 $\pm$ 3.4 days.

\begin{table}
   \begin{center}
    \caption{Fundamental stellar parameters of GJ\ 536. \label{tab:stellar_parameters}}
    \begin{tabular}{lcc}
    \hline\hline
        Parameter                       & Value                    & Ref. \\
        \hline
        RA [J2000]                      & 14:01:03.188 & 1 \\
        DEC [J2000]                     & --02:39:17.520 & 1 \\
        $\mu \alpha \cdot \cos\delta$ [mas yr$^{-1}$]& --825.463 & 1 \\
        $\mu \delta$ [mas yr$^{-1}$]    & 598.433 & 1 \\
        $\Pi$ [$mas$]                   &   95.958 $\pm$ 0.025      & 1 \\ 
        Distance [pc]                   &   10.4215 $\pm$ 0.0028   & 1\\
        $M_G$ [mag]                     &   8.7738 $\pm$ 0.0028    & 1 \\      
        $G_{BP}-G_{RP}$ [mag]           &   2.114 $\pm$ 0.005      & 1 \\  
        $m_{B}$	 [mag] & 11.177  & 2 \\
        $m_{V}$	 [mag] & 9.707   & 2 \\
        {[Fe/H]}                        &  --0.08 $\pm$ 0.09        & 3 \\
        T$_{eff}$ [K]       &  3641 $\pm$ 88            & 0 \\
        R$_{*}$ [R$_{\odot}$]            &  0.529 $\pm$ 0.022       & 0 \\
        M$_{*}$ [M$_{\odot}$]             &  0.528 $\pm$ 0.027       & 0 \\
        $\log g$ [cgs] &  4.713 $\pm$ 0.007       & 0 \\
        L$_{*}$ [L$_{\odot}$]    &  0.04437 $\pm$ 0.00092   & 0 \\
        log$_{10}$R'$_{\rm HK}$ & --5.12 $\pm$ 0.05 & 4\\
        P$_{\rm rot ~GP}$ [days] & 43.63 $\pm$ 0.85 & 0 \\
        P$_{\rm cycle}$ [days] & 3387$^{+110}_{-62}$ & 0 \\
        Rotation axis $i$ [deg] & 58$^{+16}_{-19}$ &  0\\
        Age$_{\rm Iso.}$ [Gyr]         &  9 $\pm$ 4               & 0 \\
        Age$_{\rm Gyro.}$ [Gyr]     &  4.2 $\pm$ 1.1           & 0 \\
        HZ inner edge [au] & 0.1756 $\pm$ 0.0018 & 5\\
        HZ outer edge [au] & 0.3716 $\pm$ 0.0038 & 5\\
        HZ inner period [days] & 37.1 $\pm$ 1.1 & 5\\
        HZ outer period [days] & 114.2 $\pm$ 3.4 & 5 \\

    \hline
    \end{tabular}
    \end{center}
    \textbf{References:} 0 - This work, 1 -  \citet{GaiaEDR3}, 2 - \citet{Koen2010}, 3 - \citet{Maldonado2020}, 4 - \citet{Masca2017}, 5 - Estimated following \citet{Kopparapu2014, Kopparapu2017}
\end{table}

\section{Analysis} 

We performed a global analysis of the data, always including the full set of RVs, S$_{\rm MW}$, FWHM, and photometric data. Every time series includes a zero point per instrument, a model for the cycle, and a model for the stellar rotation. The cycle model is defined as a set of sinusoidal signals, with a common period and phase for all time series, and independent amplitudes. The stellar rotation is modelled as a multi-dimensional Gaussian Process \citep{Rajpaul2015}, using the \texttt{S+LEAF} code \citep{Delisle2022} \footnote{\url{https://gitlab.unige.ch/delisle/spleaf}}. The RV data of HARPS and HARPS-N includes polynomial terms against BERV. In addition, the RV data includes a circular/Keplerian model for each planet in the model. A more detailed description of the model can be found in Appendix~\ref{append_model}.

We optimised the parameters of the models using Bayesian inference through the nested sampling \citep{Skilling2004, Skilling2006} code \texttt{Dynesty}~\citep{Speagle2020, kosopov2023}. We sampled the parameter space using random slice sampling, which is well suited for the high-dimensional spaces \citep{Handley2015a,Handley2015b} resulting from modelling several time series at once. We used a number of live points of $20 \times N_{free}$ in models with narrow priors in period, and $100 \times N_{free}$ in models with wide priors in frequency, to ensure the discoverability of the narrow frequency posteriors. 

We assessed the significance of the detection of the signals using the False inclusion probability (FIP) framework \mbox{\citep{hara2022a}}. This method uses the posterior distribution of the nested sampling run, and computes the probability of having a planet within a certain orbital frequency interval based on the relative weight of the samples within each frequency interval. The FIP framework has been demonstrated to minimise both false detections and missed detections, compared to other detection criteria such as the false alarm probability, or the Bayesian evidence. Based on an extensive set of simulations, \citet{hara2022a, Hara2024} suggested that a FIP threshold of 50\% maximises the number of detections while thresholds between 10\% and 1\%, minimise the number of mistakes (sum of false positives and false negatives). We adopt a threshold of 1\% to consider the significant detection of a signal. This method has already been shown to be effective in the detection and confirmation of exoplanets from radial velocity time series \citep{SuarezMascareno2023}

\section{Results}

\begin{figure}[!ht]
   \centering
  \includegraphics[width=9cm]{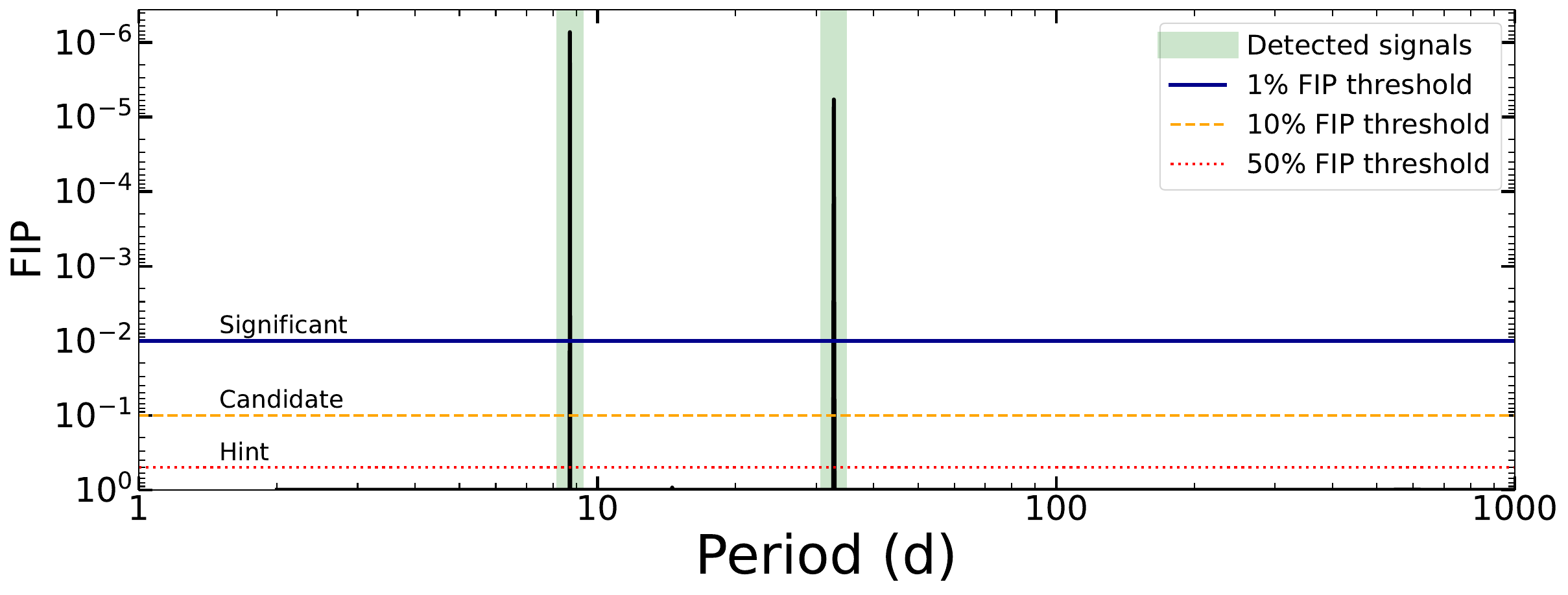}
  \caption{FIP periodogram of the full dataset. The highlighted peaks corresponds to the period of 8.7d (GJ 536 b), and 32.8d.}
  \label{fip_periodogram}
\end{figure}

We performed a blind search for planetary signals under the FIP formalism. We run a model with up to three planetary signals with wide priors and assessed the significance of the detection based on the posterior distribution of the frequency parameters. Fig.~\ref{fip_periodogram} shows the FIP periodogram of the data. The periodogram shows two signals with a FIP $<$ 0.1\%. The first signal, with a period of 8.70877 $\pm$ 0.00059 days and an RV semi-amplitude of 3.03 $\pm$ 0.15 m$\cdot$s$^{-1}$, corresponds to the planet GJ 536 b. The second, with a period of 32.761 $\pm$ 0.015 days and an RV semi-amplitude of 1.79 $\pm$ 0.22 m$\cdot$s$^{-1}$, is a new signal not previously known. There are no hints of additional signals. We cross-checked the results measuring the difference in Bayesian evidence (Ln Z) between the models with no planets, 1 planet, and 2 planets. We measure a $\Delta$ Ln Z of 53.9 favouring the 1-planet model against the no-planet model, and $\Delta$ Ln Z of 30 favouring the 2-planet model over the 1-planet model. These resuls strongly favour the 2-planet model over the 1- or 0-planet models. 

As the FIP analysis already marginalises the probability to have a planet over the activity model, it is assumed that the detected signals are likely not of stellar origin. At the very least, they are signals not accounted for the activity model, or any of the remaining components defined in the model. We measure the stellar rotation period to be 43.63 $\pm$ 0.90 days, with a timescale of coherence of the signal of 78$^{+38}_{-24}$ days. We measure the stellar cycle period to be 3390$^{+120}_{-61}$ days. 

\subsection{Presence of the signals in activity indicators}\label{presence_act}

\begin{figure*}[!ht]
   \centering
  \includegraphics[width=18cm]{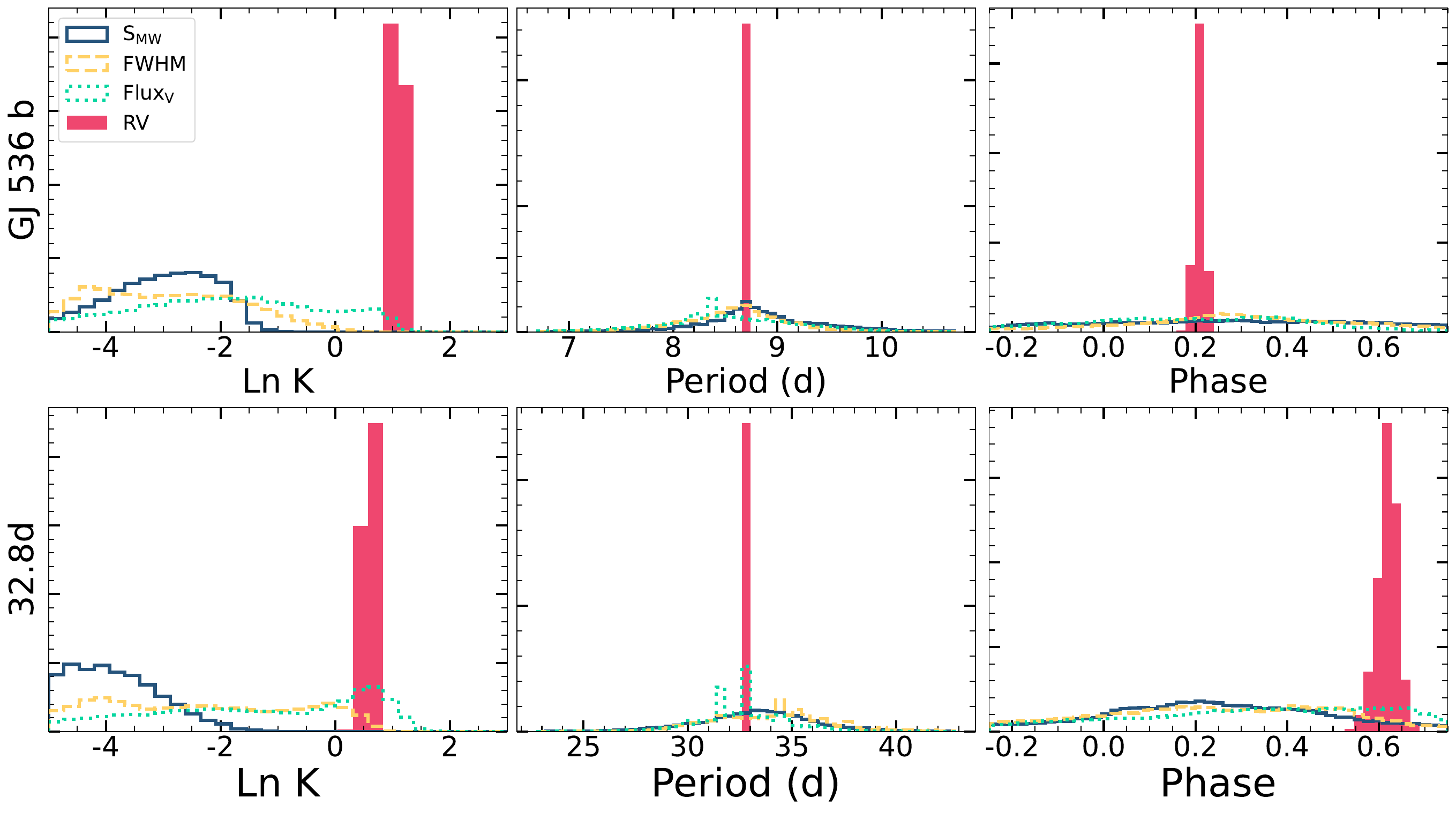}
  \caption{Comparison of the posterior of RV and activity indicators. Posterior distributions of the signal parameters of the RV signals compared with the result of performing the same model in the activity indicators.}
  \label{rv_act}
\end{figure*}

Despite having implemented an activity model in the planet search, it is possible that some activity signals might remain in the residuals of the activity model, which could lead to them being misidentified as planets. In this case, we would expect these signals to also show up in the activity indicators data. While some of their properties (amplitude, phase) would not need to match those of the RV signals, we would expect the signals to show the same periodicity. To test whether some activity signals remain at the periods of the candidate RV signals, we run a model that included two sinusoidal functions in each activity indicator with priors in commmon with the RV signals. To specifically test the regions around the detected signals, we run a guided search with normal ($\mathcal{N}$) priors centred at the determined periods, with a width of 10\% of the period. These values correspond to periods $\mathcal{N}[8.7,0.9]$ days and $\mathcal{N}[32.8,3.3]$ days. The priors for the phase and amplitude remain the same. 

Figure~\ref{rv_act} shows the posterior distributions obtained for the RV, S$_{MW}$, FWHM, and Flux data. We only obtain well-defined posteriors for the RV data. In the case of the three activity proxies, we obtain amplitudes that are 1-1.5$\sigma$ consistent with zero, and the posteriors of the period and phase are very similar to the priors. No signal appears to remain in the activity indicators with similar properties as those found in the RV data.

\subsection{Aliasing of activity signals}\label{alias}

The gaps in observability of the target stars, along with the difficulties of maintaining continuous surveys over years, typically lead to poorly sampled time series prone to aliasing problems. Poorly sampled periodic signals will not only appear at their natural frequency, but at alias frequencies due to the combination of the sampling frequency and the signal frequency. To double-check the extent in which the stellar rotation signal could cause signals that might be mistaken for any of the other two RV signals detected, we computed the window function of the data. We then produced the periodograms of the rotation signal in RV, and evaluated its main periodogram peaks and their aliases. 

The window function of our dataset has three main features: a peak at 1 day, another at 361 days, and one at 2791 days. The periodogram of the RV signal has its main activity periodicity at 43.7 days, another at 45.9 days, and then two harmonics of the 43.7d signal (21.85d and 14.55d). All those structures are presented as a forest of peaks, with the 361d aliases explaining the majority of those peaks. Figure~\ref{window_alias} shows the window function, the periodogram of the rotation-induced signals, and the position of their aliases. The periods of the detected RV signals do not coincide with any of the rotation-induced peaks, their alises, or the structures in the window function. 

\subsection{Stability of the detected signals}\label{signal_stability}

\begin{figure}[!ht]
   \centering
  \includegraphics[width=9cm]{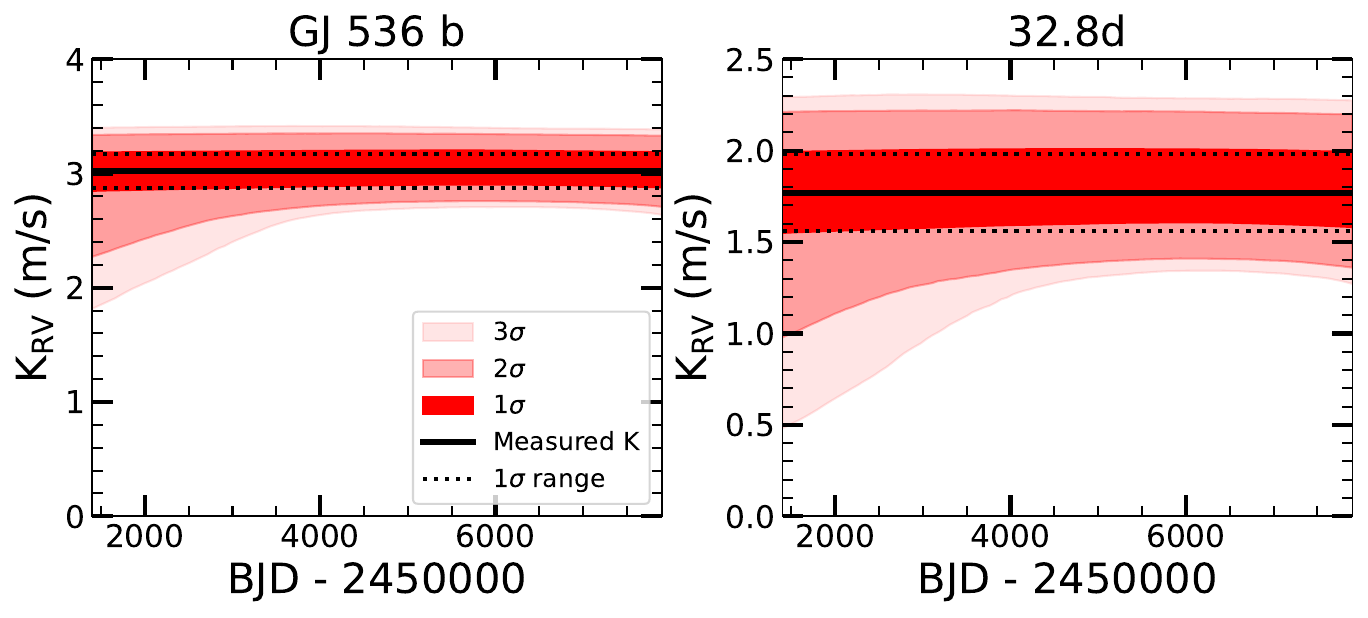}
  \includegraphics[width=9cm]{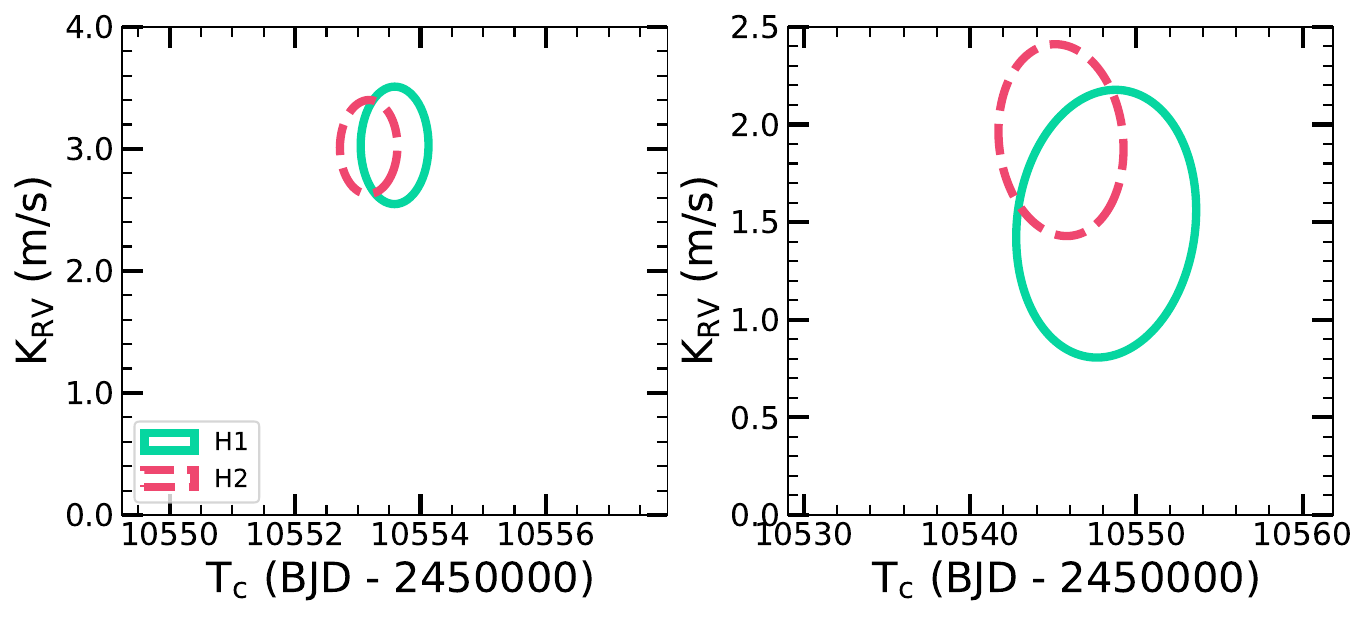}
  \caption{Stability of the signals. The top panels show the evolution of the amplitude of the apodised signals. The shaded red area shows the confidence intervals of the amplitudes of the apodised signals over time. The horizontal black lines show the median value and 1-$\sigma$ range of the static model. The bottom panels show the RV semi-amplitude vs time of inferior conjunction for the solutions obtained with the first half (H1) and second half (H2) of the dataset. The lines encapsulate the 95\% confidence interval. }
  \label{apod_amplitudes}
\end{figure}

We test the stability of the signals over time using apodised signals \citep{Gregory2016, Hara2022b}. In this test, the suspected planetary signals are multiplied by Gaussian functions with potential values of their centre ($\mu$) over the full baseline of observations, and a large range of potential widths ($\sigma$). As planetary signals are stable over time, we would expect to retrieve an undefined $\mu$ and a large $\sigma$. Retrieving very defined parameters for the Gaussian model would imply the signal is concentrated in a specific part of the baseline. We ran this test using the same narrow priors for the periods described before ($\mathcal{N}$[P, 10\% P])

Figure~\ref{apod_amplitudes}, top panels, shows the measured evolution of the amplitude, using the posterior of the apodised semi-amplitude at each point in time. The distribution of amplitudes of both signals is consistent with the values obtained in the static model. The range is wider at the beginning, coinciding with the data with the lowest observing cadence. In both cases we obtain that the centre is consistent with any point of the baseline, and the $\sigma$ is $>$ 120 000 days. 

We confirm this stability by measuring the parameters of both signals independently on the first and second half of the dataset. We split the data in two halves with equal amount of RV measurements, and ran a guided search with the same priors for the periods. Figure~\ref{apod_amplitudes}, bottom panels, shows the RV semi-amplitude versus the time of inferior conjunction of the two signals for the first and second half of the dataset, using the 95\% confidence region of the parameters. We obtain consistent parameters (within 1 $\sigma$) for both datasets. The parameters recovered in the first half of the data are less precise, which is consistent with the behaviour seen in the apodised test, due to the early data (HIRES) being less precise and having a lower cadence of observations.

\subsection{Adopted model}

\begin{figure*}[!ht]
   \centering
  \includegraphics[width=18cm]{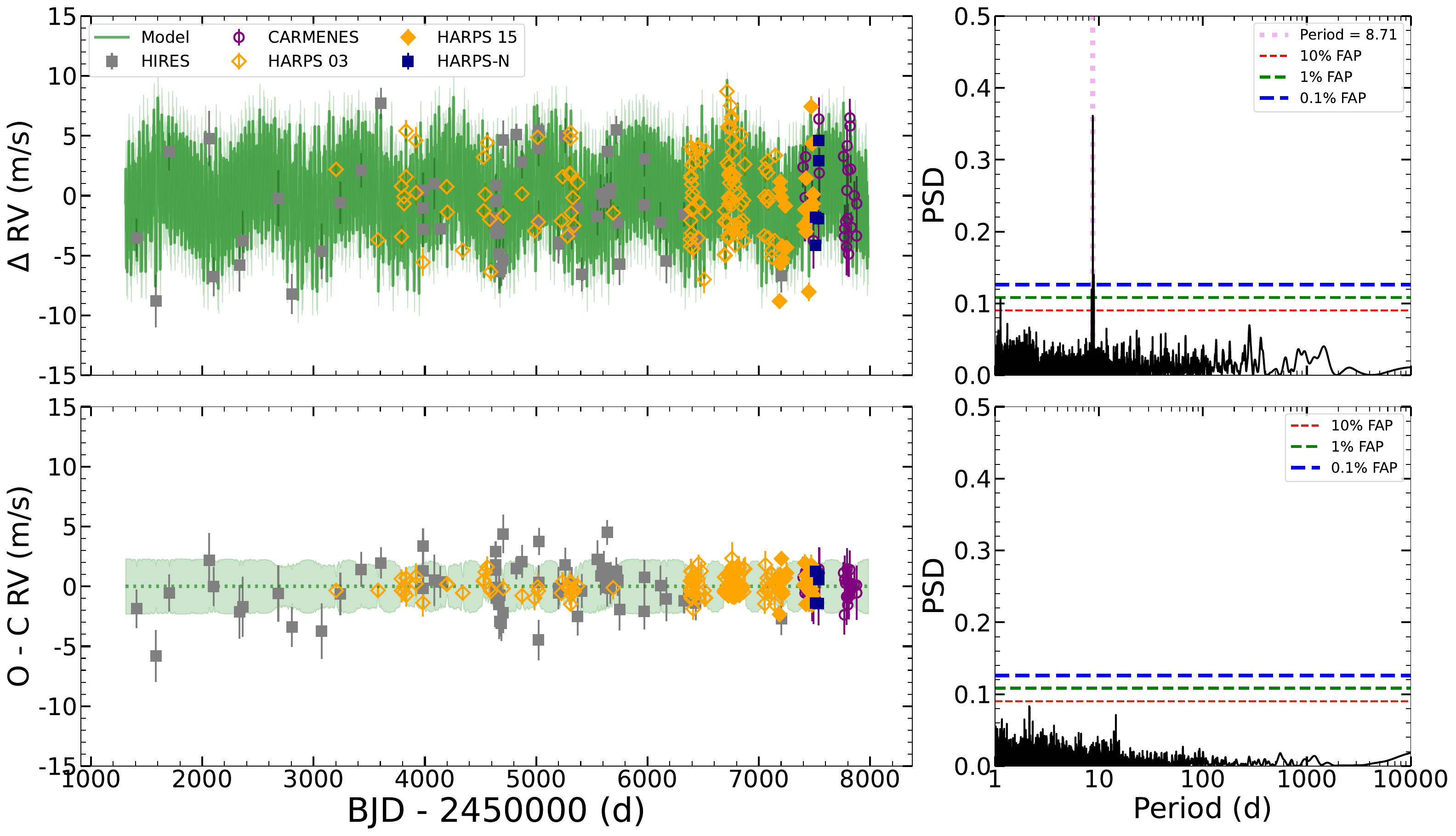}
  \caption{Adopted RV model. The top panels show the RV data (detrended from BERV), with the best model fit, along with the periodogram of the data. The bottom panels shows the residuals after the fit, along with their periodogram. Figure~\ref{adopted_act} shows the model of the activity proxies.}
  \label{adopted_rv}
\end{figure*}

\begin{figure}[!ht]
   \centering
  \includegraphics[width=9cm]{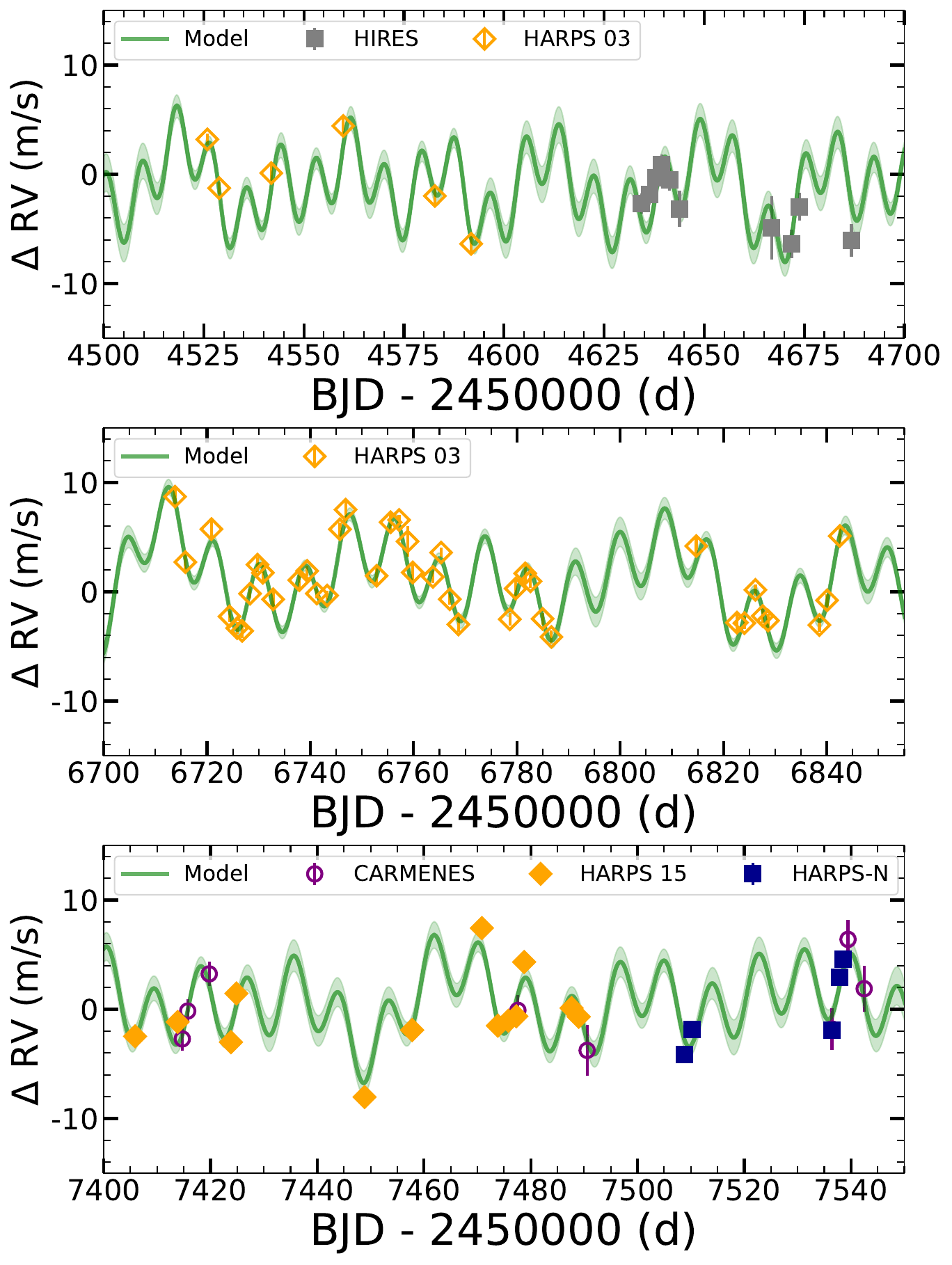}
  \caption{Zoom to selected observing campaigns. RV data of with the best model fit.}
  \label{adopted_rv_zoom}
\end{figure}

\begin{figure}[!ht]
   \centering
  \includegraphics[width=9cm]{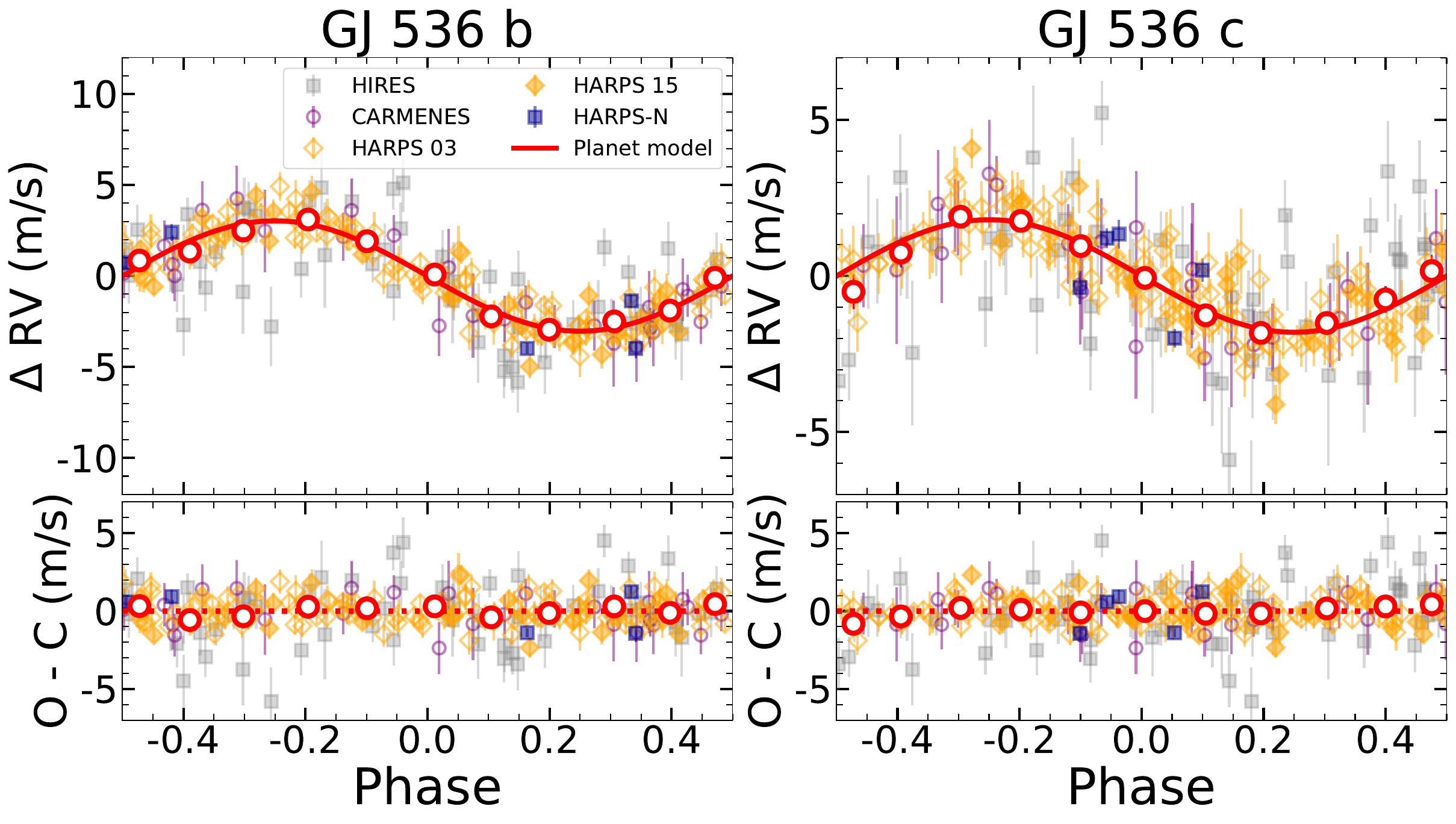}
  \caption{Phase-folded plots of the planetary-induced RV signals. RV variations induced by GJ 536 b and c with the best model fit (top panels), and the residuals after the fit (bottom panels).}
  \label{adopted_phase}
\end{figure}

Following the results of sections~\ref{presence_act},~\ref{alias} and~\ref{signal_stability}, we conclude that the signal with period 32.8 days is of planetary origin (GJ 536 c). We adopt a model that includes two sinusoidal signals to account for planets GJ 536 b and c. To establish the final parameters of these signals, we ran a model using $\mathcal{U}$ priors for the periods, centred at the known period and with a range of $\pm$ 40\% of the period. The rest of the priors for all parameters remain the same. We compared circular and Keplerian models and the results were consistent with each other. We measure upper limits to the eccentricities of 0.1 for GJ 536 b and 0.35 for GJ 536 c (95\% confidence). The rest of the parameters were consistent within 1$\sigma$. The circular model was statistically favoured, with a $\Delta$ ln Z = + 3.3 in favour of the simpler model. We adopt the circular model. The full set of priors and parameters of the adopted model is available in Appendix~\ref{append_tables}.

GJ 536 b has an orbital period of 8.70874 $\pm$ 0.00056 days and induce an RV semi-amplitude of 3.03 $\pm$ 0.15 m$\cdot$s$^{-1}$. We measure a phase of 0.208 $\pm$ 0.014, corresponding to T$_{c}$ = 10553.10 $\pm$ 0.12 (BJD -- 2450000 days). GJ 536 c has an orbital period of 32.761 $\pm$ 0.015 days, and induces an RV semi-amplitude of 1.80 $\pm$ 0.20 m$\cdot$s$^{-1}$. We measure a phase of 0.614 $\pm$ 0.029, which corresponds to T$_{c}$ = 10547.35 $\pm$ 0.96 (BJD -- 2450000 days).

We measure the stellar cycle to have a period of 3387$^{+110}_{-62}$ days, with most of its RV-induced signal manifesting at one quarter the cycle period. It induces an RMS of 0.9 m$\cdot$s$^{-1}$ in RV. We measured the stellar rotation period to be 43.63 $\pm$ 0.85 days, with a timescale of evolution of 80$^{+39}_{-25}$ days. We measure RV jitters of 0.69 $\pm$ 0.20 m$\cdot$s$^{-1}$ for the HARPS-03 data, 1.42 $\pm$ 0.40 m$\cdot$s$^{-1}$ for the HARPS-15 data, 1.57 $\pm$ 0.95 m$\cdot$s$^{-1}$ for the HARPS-N data, 0.75 $\pm$ 0.44 m$\cdot$s$^{-1}$ for the CARMENES data, and 2.00 $\pm$ 0.43 for the HIRES data. 

Figure~\ref{adopted_rv} shows the full RV dataset, along with the best-fit model, the residuals, and their respective periodograms. The model appropriately describes the data, leaving behind a residual RMS of 1.30 m$\cdot$s$^{-1}$. The periodogram of the raw dataset has a single dominant peak, at the period of GJ 536 b. No peaks at the periods of the stellar rotation, or GJ 536 c, appear. Both signals having been significantly detected in RV in our global model. This highlights the limitations of the use of periodograms to either identify signals in the data, or asses their significance. Methods that explore the full likelihood space, accounting for all potential planetary signals simultaneously, along with the variations due to stellar activity, are more efficient and robust at identifying low-amplitude planetary signals \citep{Dumusque2017,hara2022a, HaraFord2023}. Figure~\ref{adopted_rv_zoom} shows zoomed-in views of specific observing campaigns and Figure~\ref{adopted_phase} shows the phase-folded plots of the planetary-induced RV signals. Table \ref{table_adopted} shows the full set of priors, and the measured parameters of the adopted model.   

Figure~\ref{adopted_act} shows the data, periodograms, and the best-fit model, of the activity proxies used in the analyses. The cycle variations are easy to visually identify in the S$_{\rm MW}$ and FWHM time series. The periodograms both show peaks at different components of the cycle model; S$_{\rm MW}$ at P$_{\rm cyc}$ and P$_{\rm cyc}$/2, and FWHM at P$_{\rm cyc}$/2. The periodograms of both feature significant peaks at the rotation period. The periodogram of the photometry features a non-significant peak at the rotation period. Figure~\ref{adopted_act_zoom} shows zoomed-in views of specific observing campaigns.

\subsection{Estimating the inclination of the system}   

In addition to tradditional data driven GP kernels, the \texttt{S+LEAF} package includes a physics-based GP model to represent stellar activity \citep[FENRIR,][]{hara2025}, which is able to perform statistical Doppler imaging. As opposed to traditional Doppler imaging, which can retrieve a specific intensity map of the stellar surface at a given time, statistical Doppler imaging consists in retrieving the average properties of the star, in particular the projected obliquity of the stellar rotation axis and the average latitude of the active regions. The models take into account the photometric RV effect, as well as the inhibition of convective blueshift. \citet{hara2025} shows that the FENRIR model is capable of estimating the inclination of the rotation axis, and the latitudes of active regions, of the Sun, using photometry and RV data.

We ran a model that included the two planets and a FENRIR model for photometry (ASAS) and radial velocity, with the same priors for the planetary signals in RV as in the adopted model. We run spot-only, faculae-only, and combined models. In all cases, we used the \texttt{SpotsOppLatMag} version of the model, which assumes that spots appear at the two same latitutdes, symmetric with respect to the equator of the star. We also assumed that spots appear preferentially at opposite latitudes. As we do not find a significant contribution of the magnetic cycle in either photometry or RV, we did not include a magnetic cycle model. 

The spot-only model was the only one that produced consistent run-to-run results. Both faculae-only and combined models resulted in large degeneracies between the parameters. This is consistent with the expectations that activity variations in M-dwarfs is mostly spot-dominated. Using this formalism, we derive a rotation period (43.66 $\pm$ 0.69 d) and timescale of evolution (66$^{+42}_{-26}$ d) consistent with the adopted model. The resulted ativity-induced RV signal is modelled as a combination of the photometric effect (5.4 $\pm$ 1.2 m$\cdot$s$^{-1}$) and the inhibition of convection (1.75 $\pm$ 0.40 m$\cdot$s$^{-1}$). The inclination of the stellar rotation axis is estimated to be 58$^{+16}_{-19}$ degrees with respect to our line of sight, with active regions consistent with being located at latitudes of 36 $\pm$ 18 degrees. This is consistent with the spot latitude being roughly aligned with our line of sight. These results must be taken with caution, as statistical Doppler imaging is a new technique which needs further validation.

We analysed the \textit{TESS} light curve of GJ 536 looking for signs of transits, finding no such features. The lack of transits is not surprising, if we assume that the measured inclination is not too far from the real inclination of the rotaton axis. The \textit{TESS} light curve, and methods used, can be found in Appendix~\ref{tess_phot}.

\section{Discussion}

We obtained significant detections of two planetary signals in RV. These signals correspond to the already known planet GJ 536 b \citep{Masca2017}, and a new confirmed planet GJ 536 c. GJ 536 b is a planet with a minimum mass of 6.37 $\pm$ 0.38 M$_{\oplus}$, an orbital period of 8.70874 $\pm$ 0.00056 days, and an eccentricity lower than 0.1 (95\% confidence). It orbits at a distance of 0.0668 $\pm$ 0.0012 au and receives around 10 times the Earth's insolation. GJ 536 c has a minimum mass of 5.89 $\pm$ 0.70 M$_{\oplus}$, an orbital period of 32.761 $\pm$ 0.015 days, and an eccentricity lower than 0.35 (95\% confidence). It orbits at a distance of 0.1617 $\pm$ 0.0028 au, and receives an insolation slightly lower than that received by Venus, making it most likely too hot to be habitable. Table~\ref{param_planets} shows the parameters of the planets of the system of GJ 536, using the adopted (circular) model. Figure~\ref{planets_context} shows the position of the planets in the context of known low-mass exoplanets, as a function of orbital period and stellar insolation. GJ 536 c is among the group of lowest-mass exoplanets known at moderate orbital periods, and one with comparatively low insolation. 

Planets GJ 536 b and c have very similar minimum masses (6.37 $\pm$ 0.38 M$_{\oplus}$ and 5.89 $\pm$ 0.70 M$_{\oplus}$, respectively). Their nature would be extremely dependendent of the inclination of their orbits with respect to the plane of the sky. For large inclinations (i $>$ 45$^{\circ}$), they will most likely be massive super-Earths, water-worlds, or mini-Neptunes. With inclinations $<$ 45$^{\circ}$, they will most likely be mini-Neptunes or even Neptune-like planets. For them to be gas giants, it would be neccessary that the inclination of the orbital plane is $<$ 5$^{\circ}$. If we assume the inclination of the rotation axis derived using the FENRIR model, and the orbital plane to be coplanar with the rotation axis, their masses would be 7.6$^{+2.3}_{-1.0}$ M$_{\oplus}$ and 7.1 $^{+2.2}_{-1.2}$ M$_{\oplus}$, respectively, making them most likely super-Earths or water-worlds. This estimation, however, needs to be taken with caution, given the number of assumptions required for the computation.

Our current measure of the mass of GJ 536 b (6.37 $\pm$ 0.38 M$_{\oplus}$) is slightly larger than the measurement of \citet{Masca2017} (5.36 $\pm$ 0.69 M$_{\oplus}$). The difference comes from a larger measured RV amplitude of 3.03 $\pm$ 0.15 m$\cdot$s$^{-1}$, versus the original 2.60 $\pm$ 0.33 m$\cdot$s$^{-1}$. For the current analysis, we used a larger dataset, an RV extraction method for the HARPS data more suited for M-dwarf spectra, and a more sophisticated model for stellar activity. The limitations of the original study likely lead to a slight underestimation of the RV amplitude.

\subsection{Direct imaging characterisation}

\begin{figure}[!ht]
   \centering
  \includegraphics[width=9cm]{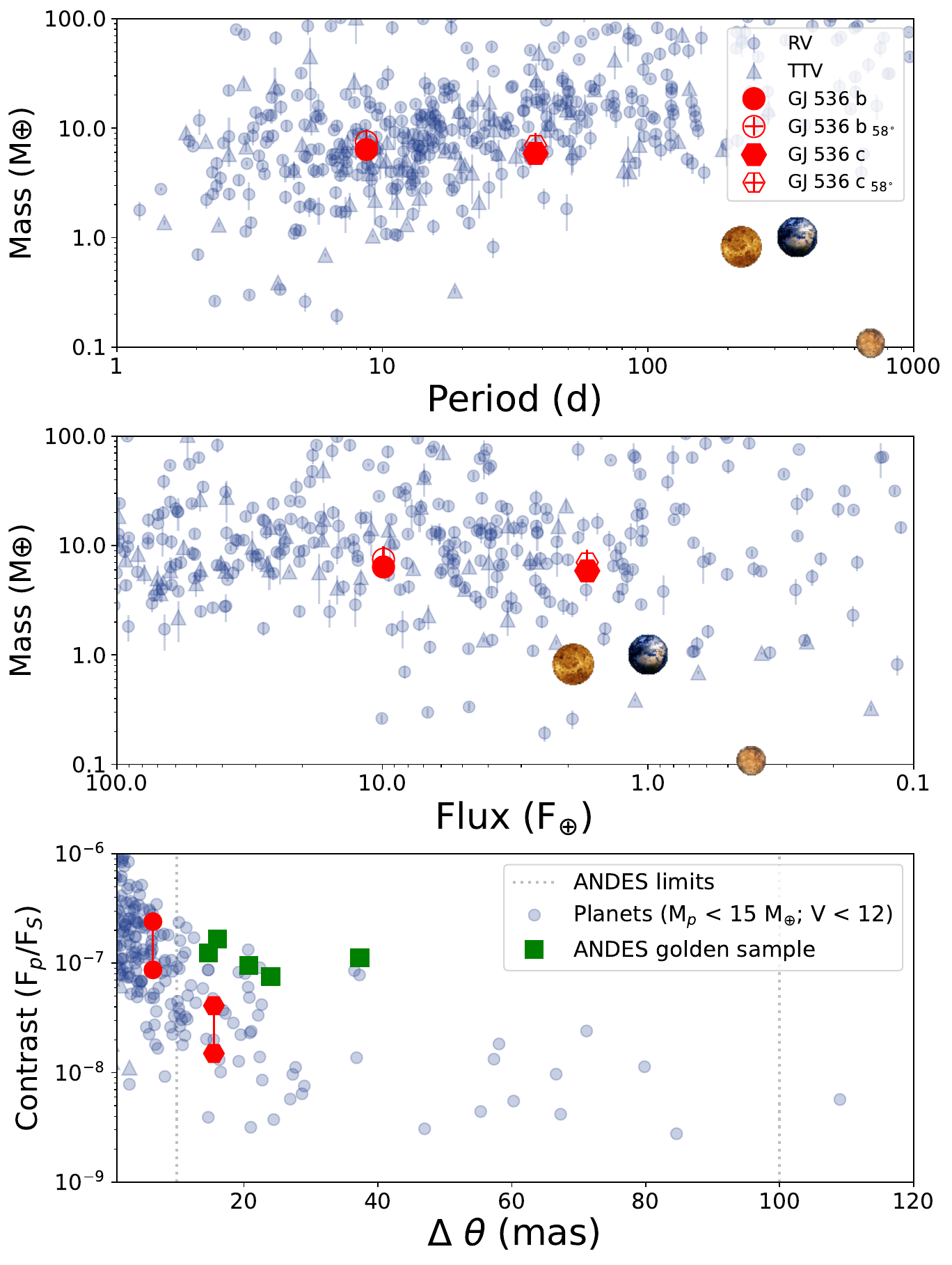}
  \caption{GJ 536 b and c in context. The upper panel shows the masses of known planets, for masses below 100 M$_{\oplus}$, and periods shorter than 1000 days, with the planets of the system of GJ 536 highlighted, and the solar system planets as reference. The middle panel shows the same, but as a function of insolation. The lower panel shows the contrast in reflected light of low-mass planets orbiting bright stars (m$_{V}$ < 12), compared to their angular separation to their parent star. The dotted vertical lines indicate the limits in which ANDES wil be able to observe \citep{Palle2025}. The red symbols for the planets of GJ 536 show the potential range of contrasts. The green squares show the position of the ANDES golden sample.}
  \label{planets_context}
\end{figure}

The angular separation of GJ 536 b and c is of 6.41 $\pm$ 0.11 mas, and 15.51 $\pm$ 0.26 mas, respectively. Assuming their radii are between 1.5 and 2.5 R$_{\oplus}$ (as most planets of similar masses), and albedos of 0.3, we estimate a F$_{p}$/F$_{S}$ contrast in reflected light between 8.7 $\cdot$ 10$^{-8}$ and 2.4 $\cdot$ 10$^{-7}$ for planet b, and between 1.8 $\cdot$ 10$^{-8}$ and 1.5 $\cdot$ 10$^{-8}$ for planet c. While GJ 536 b is too close to its star, GJ 536 c is a good candidate for direct imaging with future giant telescopes, such as the ELT with ANDES \citet{Marconi2022}, and missions such as LIFE \citet{Quanz2022}. Figure~\ref{planets_context}, lower panel, shows their position in comparison with other low-mass planets (m$_{p}$ < 100 M$_{\oplus}$) orbiting bright stars (V < 12). GJ 536 c is among the 20-30 most favourable targets for atmosphere characterization via direct imaging, depending on the specific radius assumption and magnitude cut.

\begin{table} 
   \begin{center}
   \caption{Parameters of the planets of the system of GJ 536, using the adopted (circular) model.  \label{param_planets}}
   \begin{tabular}[centre]{l c}
   \hline
   Parameter  & Value \\ \hline
   \textbf{GJ 536 b} \\
   $T_{0}$ -- 2450000 [d] & 10553.10 $\pm$ 0.12 \\
   $P_{\rm orb}$ [d] & 8.70874 $\pm$ 0.00056  \\
   $m_p$ sin $i$ [M$_{\oplus}$] & 6.37 $\pm$ 0.38 \\
   $m_{p~58^{\circ}}$ [M$_{\oplus}$] & 7.6$^{+2.3}_{-1.0}$ \\
   $a$ [au] &  0.0668 $\pm$ 0.0012 \\
   $\Delta \theta$ [mas] & 6.41 $\pm$ 0.11 \\
   F$_{p}$/F$_{S}$$^{1}$ & 8.7 $\cdot$ 10$^{-8}$ to 2.4 $\cdot$ 10$^{-7}$ \\
   $e$ & 0 (fixed)\\
   Incident flux [$S_{\oplus}$] & 9.90 $\pm$ 0.41  \\
   T$_{\rm eq~A = 0.3}$ [K] & 451 $\pm$ 15  \\ 
   K [m$\cdot$s$^{-1}$]  & 3.03 $\pm$ 0.15  \\ \\
   \textbf{GJ 536 c} \\
   $T_{0}$ -- 2450000 [d] & 10547.35 $\pm$ 0.96  \\
   $P_{\rm orb}$ [d] & 32.761 $\pm$ 0.015 \\
   $m_p$ sin $i$ [M$_{\oplus}$] &  5.89 $\pm$ 0.70 \\
   $m_{p~58^{\circ}}$ [M$_{\oplus}$] & 7.1 $^{+2.2}_{-1.2}$ \\
   $a$ [au] &   0.1617 $\pm$ 0.0028\\
   $\Delta \theta$ [mas] & 15.51 $\pm$ 0.26 \\
   F$_{p}$/F$_{S}$$^{1}$ & 1.5 $\cdot$ 10$^{-8}$ to 4.1 $\cdot$ 10$^{-8}$\\
   $e$ & 0 (fixed) \\
   Incident flux [$S_{\oplus}$] &  1.692 $\pm$ 0.069 \\
   T$_{\rm eq~A = 0.3}$ [K] & 290.5 $\pm$ 9.5 \\ 
   K [m$\cdot$s$^{-1}$]  & 1.80 $\pm$ 0.20 \\
   \hline
   \end{tabular}       
   \end{center}
   $^{1}$ Reflected light. Albedo = 0.3.
   \end{table}

\subsection{Limits on additional companions}

\begin{figure}[!ht]
   \centering
  \includegraphics[width=9cm]{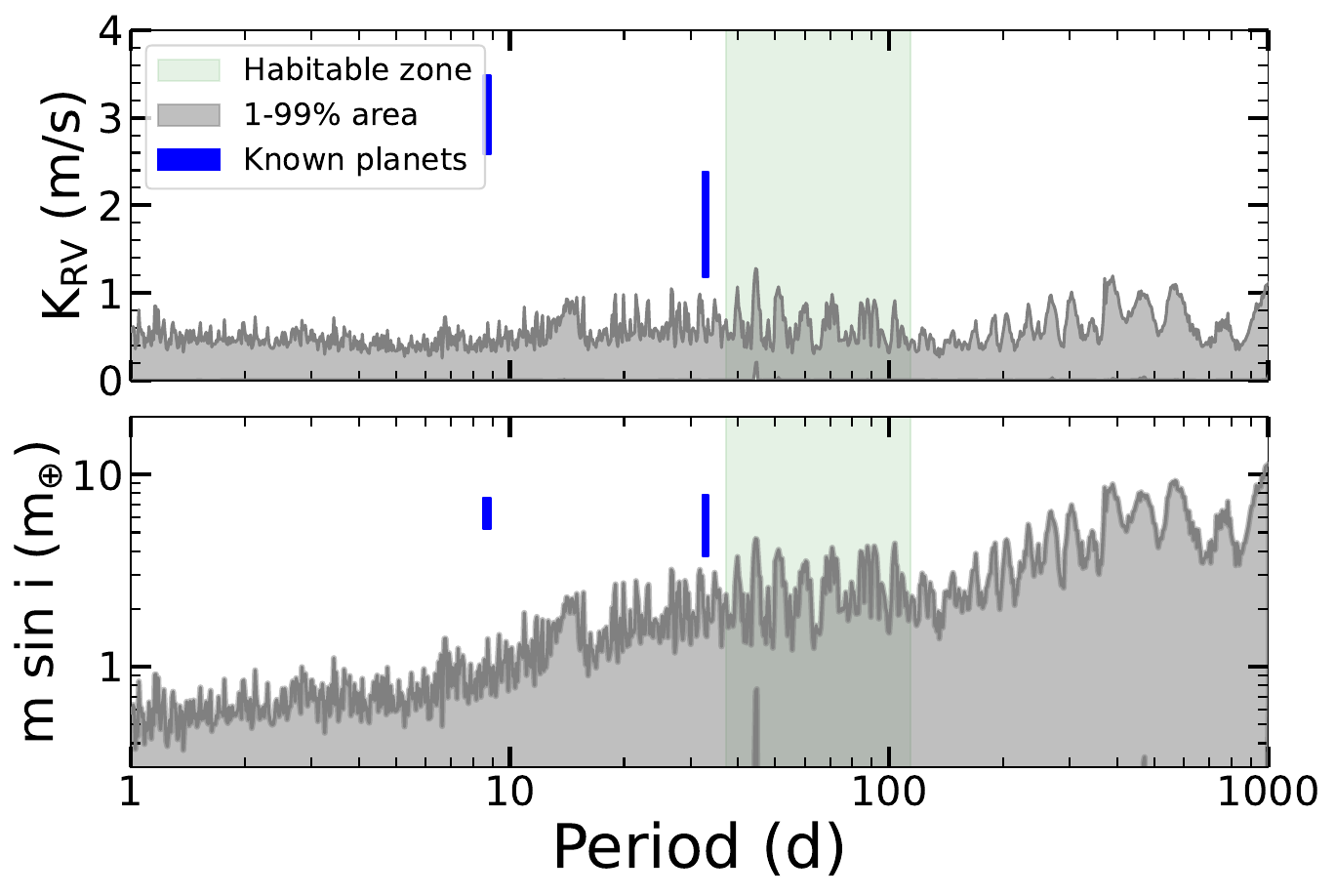}
  \caption{Compatibility limits. The upper panel shows the RV amplitude limit (1-99\% area) as a function of orbital period. The lower panels show the same, but for planetary masses. The blue rectangles show the 3$\sigma$ amplitude/mass range of the detected planets.}
  \label{fig_det_lims}
\end{figure}

Using the results from the adopted model, we measured the compatibility limits for an additional planetary signal at a wide range of orbital periods. We froze most of the parameters of the model (trends, cycle, GP, planets) and left free only the white noise and zero point RV components. We included a third sinusoid in the model. We built a grid of 1000 bins over periods 1 - 1000 days with equal width in log$_{10}$ space. We ran this model with a narrow prior for the period within each bin. From the posterior distribution of each try, we computed the 1\% and 99\% limits in RV amplitude. Figure~\ref{fig_det_lims} shows the result of this exercise. The shaded area encapsulates this range. Any potential signal left in the data should show as a deviation from zero in RV amplitude (or mass). We found that, for periods shorter than ~10 days, we can exclude the presence of any additional signal with an amplitude larger than 50 cm$\cdot$s$^{-1}$ (minimum masses 0.6 -- 1.5 M$_{\oplus}$). Within the habitable zone, we can exclude, for the most part, the presence of signals with amplitudes between 0.5 and 1.2 m$\cdot$s$^{-1}$ ($m$ sin $i$ 2 -- 4 M$_{\oplus}$). At periods between the edge of the habitable zone (114 d) and 1000 days, we can exclude the presence of additional signals beyond 0.5 -- 1.2 m$\cdot$s$^{-1}$ ($m$ sin $i$ of 2 -- 10 M$_{\oplus}$). 

There is one period bin in which the measured RV amplitude is not consistent with zero. At 44 days we measure an RV semi-amplitude of 76 $\pm$ 22 cm$\cdot$s$^{-1}$. This signal did not appear in our global model optimisation and the period is consistent with the measured stellar rotation. We believe it to be a leftover of stellar rotation due to having fixed the activity parameters rather than optimising them. 

\subsection{Stellar activity of GJ 536}

Along with the planetary system, we characterised the activity variations of GJ 536. We analysed the time series of the V-mag flux, Ca II H\&K lines, the RV, and the FWHM of the stellar lines simultaneously with a model that combined a cycle component and Gaussian processes. 

\begin{figure}[!ht]
   \centering
  \includegraphics[width=9cm]{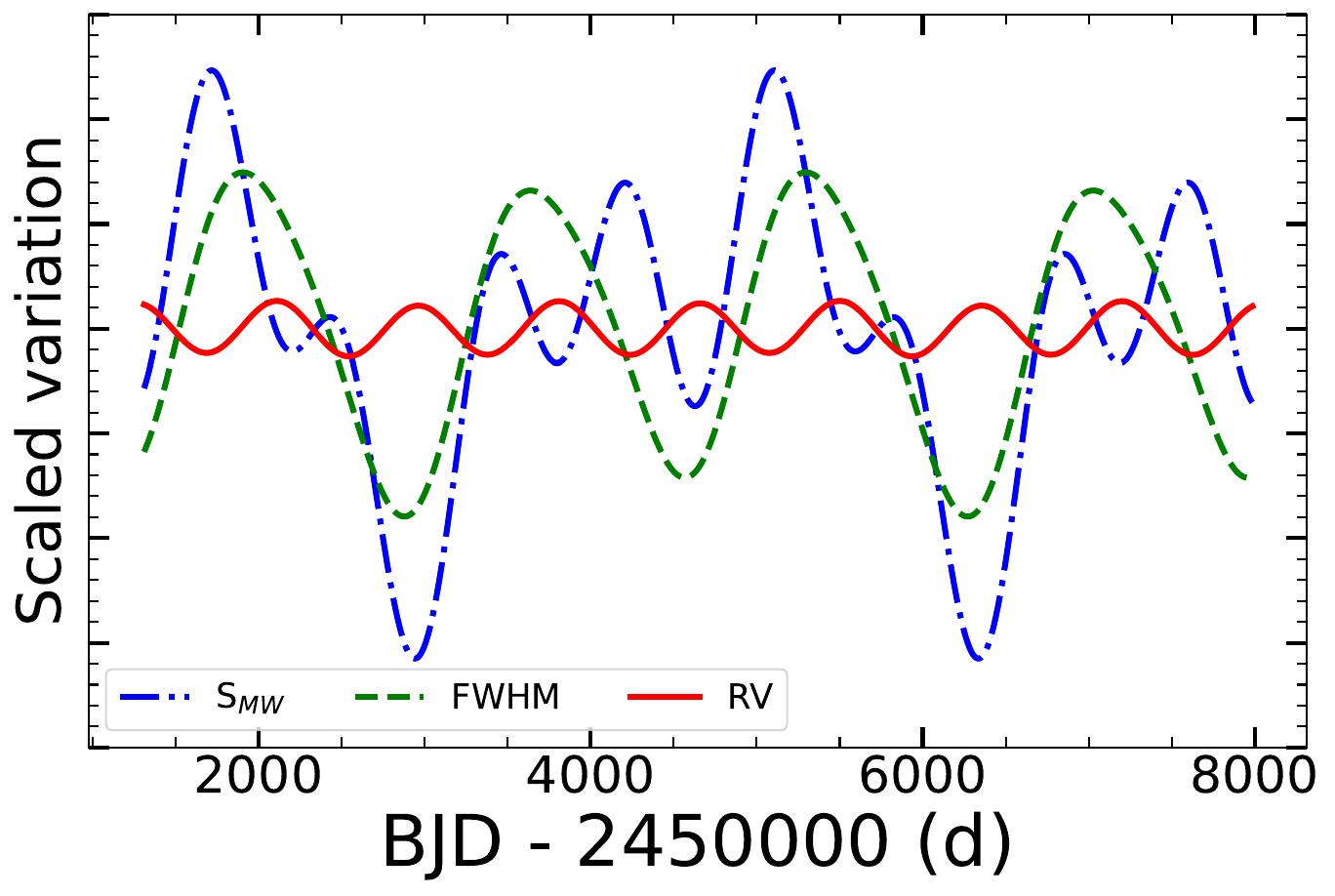}
  \caption{Comparison of the cycle variations. Cycle model in S$_{\rm MW}$, FWHM, and RV, showing the relationship between the components. The models have been scaled as if all original time series had the same RMS.}
  \label{cycle_models}
\end{figure}

We measured a cycle period of 3387$^{+110}_{-62}$ days (9.27$^{+0.30}_{-0.17}$ years). The period is in the ballpark of the longer period reported by \citet{IbanezBustos2025}. The shape of the magnetic cycle is significantly different from a sinusoidal. Instead, it is better described by a combination of several, at the natural period and its harmonics. This shape of the cycle explains the period measurements at $\sim$ 1700 days and $\sim$ 850 days estimated in previous works \citep{Masca2017,IbanezBustos2025}. We found that the main variability of the cycle did not manifest at the same period for all time series. The S$_{\rm MW}$ data showed variations at the main period, half, and one quarter. The FWHM data only at half the period, and the RV data only at one quarter. Figure~\ref{cycle_models} shows the comparison between the cycle models in the three time series. The cycle creates a variation in S$_{\rm MW}$ of 0.05 RMS (total RMS 0.1), of 3.2 m$\cdot$s$^{-1}$ RMS in FWHM (total RMS 6.9 m$\cdot$s$^{-1}$), and of 1.0 m$\cdot$s$^{-1}$ RMS in RV (total RMS  3.5 m$\cdot$s$^{-1}$).

Fully disentangling whether we see a complex cycle, which different proxies track in different ways, or the real cycle is shorter ($\sim$ 1700 days) and the S$_{\rm MW}$ index data is affected by some additional source of noise is difficult with the data at hand. We compared models with longer and complex cycles, against shorter and simpler cycles, and tried to evaluate the best using the difference in Bayesian evidence ($\Delta$ lnZ). The model featured in the analysis (long period with power at P$_{\rm cyc}$/2 and P$_{\rm cyc}$/4) was favoured if we used all time series simultaneously (RV, S$_{\rm MW}$, FWHM, and Flux$_{\rm V}$; $\Delta$ lnZ > 3.5 over the second-best model). Removing the RVs from the equation, the favoured period remains consistent, although simpler shapes were favoured (P$_{\rm cyc}$ and P$_{\rm cyc}$/2 only).

Using GP regression, we measured the stellar rotation period to be 43.63 $\pm$ 0.85 days, with a timescale of evolution of 80$^{+39}_{-26}$ days. The measured period is consistent with previous measurements \citep{Masca2015,Masca2017}, with a timescale of coherence of about two times that period. The rotation signal creates a variation in photometry of 3.27 ppt RMS, of 0.05 RMS in S$_{\rm MW}$, of 4.2 m$\cdot$s$^{-1}$ RMS in FWHM, and of 1.8 m$\cdot$s$^{-1}$ RMS in RV. 

Analysing the posterior distribution of the GP parameters, we find that the rotation variations in Ca II H\&K and photometry are anti-correlated. The variations in Ca II H\&K and FWHM are positively correlated. The variations in RV are positively correlated with the variations in Ca II H\&K and with their gradient, and negatively with those of photometry and their gradient. The RV component linked to the gradient of the flux dominates over that linear with the flux.  This behaviour is consistent with a spot-dominated stellar surface in which plages surround the spots. Lower fluxes (more spots) come together with higher Ca II H\&K fluxes. The stellar RV variations are consistent with being produced by both inhibition of convection and the photometric effect. This is reinforced by the results coming from the FENRIR model.

\section{Conclusions}

We revisited the system GJ 536 using all available spectroscopic data from HARPS, HARPS-N, CARMENES, and HIRES, and updated RV-extraction, and analysis methods. We performed a joint model that combined RV with information of different activity proxies, into the multi-dimensional Gaussian process framework. 

Our analysis provided updated parameters of the known planet GJ 536 b, and the discovery of a new planet (GJ 536 c). GJ 536 b has an orbital period of 8.70874 $\pm$ 0.00056 days, a minimum mass of 6.37 $\pm$ 0.38 M$_{\oplus}$, and receives $\sim$ 10 times the Earth's insolation. GJ 536 c has an orbital period of 32.761 $\pm$ 0.015 days, a minimum mass of 5.89 $\pm$ 0.70 M$_{\oplus}$, and receives an insolation slightly lower than that received by Venus. Baed on statistical Doppler imaging, we estimated an inclination of the rotation axis of 58$^{+16}_{-19}$ degrees. If we assume the orbital plane to be coplanar with the rotation axis, their masses would be 7.6$^{+2.3}_{-1.0}$ M$_{\oplus}$ and 7.1$^{+2.2}_{-1.2}$ M$_{\oplus}$, respectively. Given its projected angular separation of 15.5 mas, expected planet to star contrast of 1.8 $\cdot$ 10$^{-8}$ to 3.0 $\cdot$ 10$^{-8}$, and the brightness of the star, GJ 536 c is among the select group of low-mass exoplanets amenable for atmospheric characterization using its reflected light. 

In addition to the analysis of the planetary system, we analysed the activity of the star. GJ 536 has a 9.27$^{+0.30}_{-0.17}$ years activity cycle, detectable in the variations of the Ca II H\&K fluxes, the FWHM of the stellar lines (at half the period), and the RV (at one quarter the period). The star rotates every 43.63 $\pm$ 0.81 days, and the induced signals are consistent with a spot-dominated surface in which plages and spots coexist. 

The analysis of GJ 536 highlights the advances in data extraction and activity mitigation that have been made over the past years. These advances open the door to detecting planets that may have previously escaped detection, revealing new worlds hidden within archival data.

\section*{Data availability}

The time series data used in this paper is available in electronic form at the CDS via anonymous ftp to cdsarc.u-strasbg.fr (130.79.128.5) or via http://cdsweb.u-strasbg.fr/cgi-bin/qcat?J/A+A/. 

\begin{acknowledgements}

A.S.M., J.I.G.H., and R.R. acknowledge financial support from the Spanish Ministry of Science and Innovation (MICINN) project PID2020-117493GB-I00 and from the Government of the Canary Islands project ProID2020010129. 

CdB acknowledges support from the Agencia Estatal de Investigación del Ministerio de Ciencia, Innovación y Universidades (MCIU/AEI) under grant WEAVE: EXPLORING THE COSMIC ORIGINAL SYMPHONY, FROM STARS TO GALAXY CLUSTERS and the European Regional Development Fund (ERDF) with reference PID2023-153342NB-I00/10.13039/501100011033, as well as from a Beatriz Galindo Senior Fellowship (BG22/00166) from the MICIU. The University of La Laguna (ULL) and the Department of Economy, Knowledge, and Employment of the Government of the Canary Islands are also gratefully acknowledged for the support provided to CdB (2024/347).

The project that gave rise to these results received the support of a fellowship from the ”la Caixa” Foundation (ID 100010434). The fellowship code is LCF/BQ/DI23/11990071.

NN acknowledges funding from Light Bridges for the Doctoral Thesis "Habitable Earth-like planets with ESPRESSO and NIRPS", in cooperation with the Instituto de Astrofísica de Canarias, and the use of Indefeasible Computer Rights (ICR) being commissioned at the ASTRO POC project in the Island of Tenerife, Canary Islands (Spain). The ICR-ASTRONOMY used for his research was provided by Light Bridges in cooperation with Hewlett Packard Enterprise (HPE).

Part of this research was carried out at the Jet Propulsion Laboratory, California Institute of Technology, under a contract with the National Aeronautics and Space Administration (NASA)

\newline
This work is based on data obtained via the  HARPS public database at the European Southern Observatory (ESO). We are grateful to all the observers of the following ESO projects, whose data we are using: 072.C-0488, 085.C-0019, 183.C-0972, and 191.C-087. We are grateful to the crews at the ESO observatories of Paranal and La Silla. 

This research has made extensive use of the SIMBAD database, operated at CDS, Strasbourg, France, and NASA's Astrophysics Data System. 

This research has made use of the NASA Exoplanet Archive, which is operated by the California Institute of Technology, under contract with the National Aeronautics and Space Administration under the Exoplanet Exploration Program.

\newline
The manuscript was written using \texttt{VS Code}. 
Main analysis performed in \texttt{Python3} \citep{Python3} running on \texttt{Ubuntu} \citep{Ubuntu} systems and \texttt{MS. Windows} running the \texttt{Windows subsystem for Linux (WLS)}.
Extensive usage of \texttt{Numpy} \citep{Numpy}.
Extensive usage of \texttt{Scipy} \citep{Scipy}.
All figures built with \texttt{Matplotlib} \citep{Matplotlib}.
The bulk of the analysis was performed on desktop PC with an AMD Ryzen$^{\rm TM}$ 9 9950X (16 cores, 32 threads, 3.5--4.7 GHz), provided by ASM, and a server hosting 2x AMD Epyc$^{\rm TM}$ 7663 (56 cores, 112 threads, per cpu), provided by the SUBSTELLAR ERC AdG.

\end{acknowledgements}

%
%
\bibliography{biblio_gj536}

\begin{appendix}

\clearpage

\onecolumn

\section{Correlations between RV and activity indicators} \label{append_correl}

\begin{figure}[!ht]
   \centering
  \includegraphics[width=18cm]{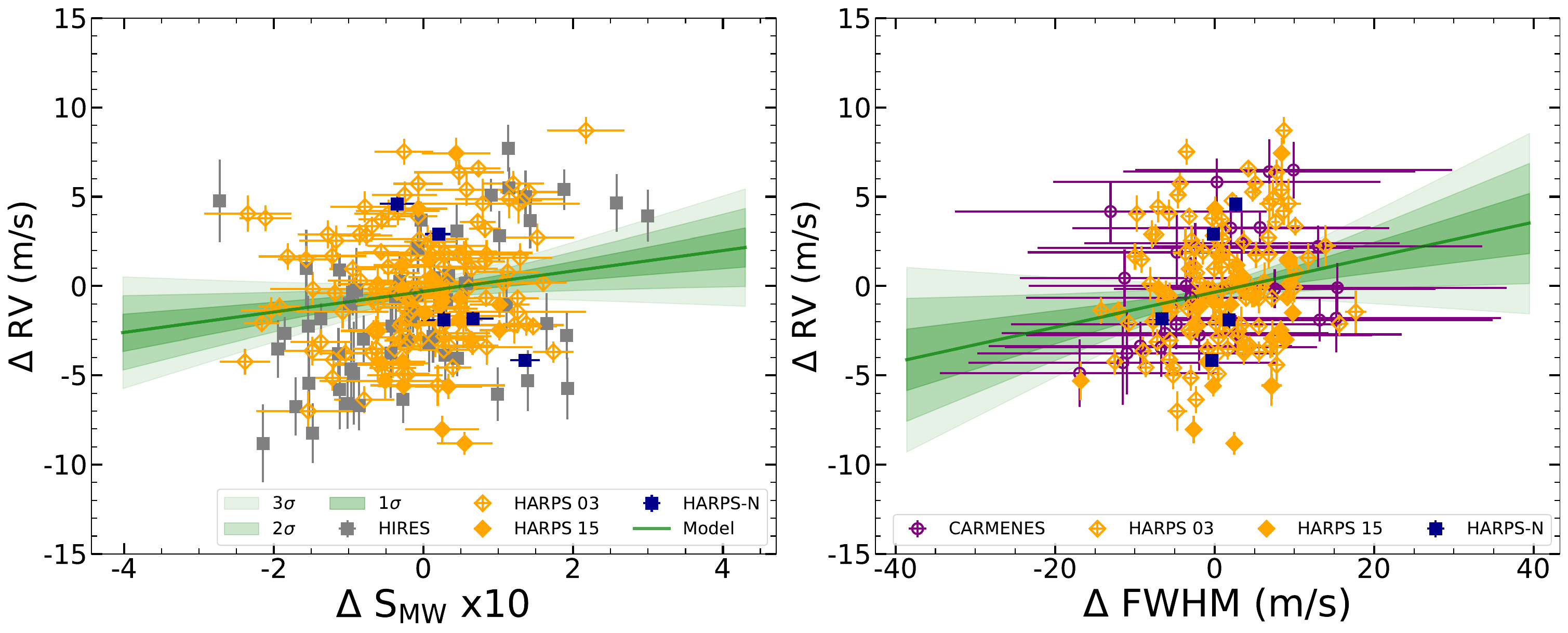}
  \caption{RV against $S$ index $\times$ 10 (left panel) and FWHM (right panel) for GJ 536. The green line, and shaded areas, show the best fit to the data and the confidence intervals of the models.}
  \label{correlation_act_rv}
\end{figure}

We studied the level of correlation between RV and the used spectroscopic activity indicators. Fig.~\ref{correlation_act_rv} shows th RV measurements against $S$ index and FWHM. We computed the level of correlation using Spearman's rank correlation coefficient \citep{Spearman1904}. As the correlation coefficient does not take uncertainties into account, we performed 10~000 independent measures by taking random values from the distribution given by the error bars of the data, and computed the median result and its standard deviation. We measured a correlation coefficient of 0.216 $\pm$ 0.032 for the RV vs $S$ index and 0.147 $\pm$ 0.042 for the RV vs FWHM. These values imply a weak correlation between the RV and the activity indicators. In addition, we modelled the relationship between these variables as a first-order polynomial. We measured a RV vs $S$ index slope of 0.57 $\pm$ 0.25 m$\cdot$s$^{-1}$ and a RV vs FWHM slope of 0.098 $\pm$ 0.043. Both slopes are 3$\sigma$ consistent with zero, supporting the weak level of correlation.

\section{Model description} \label{append_model}

We performed global analyses of the data that included photometry, FWHM, S$_{MW}$ index, and RV. Every time series includes a zero point per instrument, and models for the stellar cycle and rotation. 
The stellar rotation is modelled as a Gaussian Process. The HARPS FWHM data included a linear term to account for the focus drift of the instrument. The HARPS RV data, in addition, includes a polynomial term against the BERV. On top of them, the RV data includes a circular/Keplerian model for each planet in the model. To avoid numerical issues, we scaled up the S$_{MW}$ index as S$_{MW}$ index $\times$ 10.

The full model is defined as:   

\begin{equation} \label{eq_full_model}
    \begin{split}
    \Delta ~Flux = & ~V0 + Cycle + Rot ~,\\
    \Delta ~S_{MW} \times 10 = & ~V0 + Cycle + Rot~, \\
    \Delta ~FWHM = & ~V0 + f_{Focus} + Cycle + Rot~, \\
    \Delta ~RV = & ~V0 + f_{BERV} + Cycle + Rot + Planets~,\\
\end{split}
\end{equation}

\noindent where V0 represents the zero-point of each time series, with priors $\mathcal{N}[0,10]$ ppt for the photometric data,  $\mathcal{N}[0,10]$ for the S$_{MW}$ $\times$ 10, $\mathcal{N}[0,10]$ m$\cdot$s$^{-1}$ for the FWHM, and $\mathcal{N}[0,10]$ m$\cdot$s$^{-1}$ for the RVs. The rest of the components are described in the following subsections.

\subsection{Cycle model}

Based on the results of a S$_{MW}$-only model, the cycle component is defined as: 

\begin{equation} \label{eq_full_model}
    \begin{split}
    \Delta ~y = &~ -A_{1} \cdot sin(2\pi (t - t_{1})/P_{\rm cyc})\\
    &~ -A_{2} \cdot sin(4\pi (t - t_{2})/P_{\rm cyc})  \\
    &~ -A_{3} \cdot sin(8\pi (t - t_{3})/P_{\rm cyc})  
\end{split}
\end{equation}

\noindent with:

\begin{equation} \label{eq_full_model_par}
    \begin{split}
    &t_{1} = t_{0} + P_{\rm cyc} \cdot (\phi_{1} - 1) \\
    &t_{2} = t_{0} + P_{\rm cyc} \cdot (\phi_{2} - 1) / 2\\
    &t_{3} = t_{0} + P_{\rm cyc} \cdot (\phi_{3} - 1) / 4
\end{split}
\end{equation}

\noindent with $t_{0} =  7880$ (BJD -- 2\,450\,000) being the integer date of the last RV observation. The cycle has independent amplitudes for photometry, S$_{MW}$  $\times$ 10, FWHM, and RV. The period of the cycle has a prior Ln P $\mathcal{U} [7, 8.7]$ (1100 -- 6000 days).The phases have priors $\mathcal{U} [-0.5 , 0.5]$. The amplitudes in S$_{MW}$  $\times$ 10 are restricted to be positive ($\mathcal{U} [0 , 2]$), while the amplitudes in Photometry, FWHM and RV can be either positive or negative, to account for opposition of phase ($\mathcal{N} [0 , 10]$ in their respective units for all of them).

\subsection{Stellar rotation model}

To model the rotation we opted to work within the multidimensional Gaussian Processes (GP) framework \citep{Rajpaul2015}, which is based on the assumption that there exists an underlying function governing the behaviour of the stellar activity in all time series, which we denote $G(t)$. $G(t)$ manifests in each time series (Photometry, RV, etc.) as a linear combination of itself and its gradient, $G'(t)$, with a set of amplitudes for each time series, following the idea of the $FF'$ formalism \citep{Aigrain2012}. We used the \texttt{S+LEAF} code \citep{Delisle2022} \footnote{\url{https://gitlab.unige.ch/delisle/spleaf}}, which extends the formalism of semi-separable matrices introduced with \texttt{celerite} \citep{Foreman-Mackey2017} to allow for fast evaluation of GP models even in the case of large datasets.  The \texttt{S+LEAF} code supports a wide variety of GP kernels with fairly different properties. We opted for a combination of two stochastic harmonic oscillators (SHO), one with a period equal to the rotation period, and the second at half the rotation period. The kernel is defined as: 

\begin{equation} \label{act_model}
    \begin{split}
    \fontsize{8}{11}\selectfont
     k(\tau, P_{\rm rot}, L) = k_{SHO ~\rm 1}(\tau, \alpha_{1}, P_{1}, Q_{1}) + k_{SHO ~\rm 2}(\tau, \alpha_{2}, P_{2}, Q_{2}) \\
     + (\sigma^2 (t) + \sigma^2_{j}) \cdot \delta_{\tau}
     \end{split}
\end{equation}

\noindent where $k$ denotes the GP term, $\alpha_{i}$ is the standard deviation of the process,  $P_{i}$ is the period of the oscillator, related to the rotation period, and $Q_{i}$ is the quality factor of the oscillator, related to the timescale of evolution $L$. These parameters are defined as: 

\begin{equation} \label{eq_params}
\begin{split}
&\alpha_{1} = 1 ~; P_{1} = P_{\rm rot} ~; Q_{1} = {\pi {L}\over{P_{\rm rot}}} \\
&\alpha_{2} = \beta ~; P_{2} = 0.5 \cdot P_{\rm rot} ~; Q_{2} = {2\pi {L}\over{P_{\rm rot}}} \\
\end{split}
\end{equation}

\noindent Equation~\ref{act_model} also includes a term of uncorrelated noise ($\sigma$), independent for every instrument, added quadratically to the diagonal of the covariance matrix to account for all unmodelled sources of variation, such as uncorrected activity, instrumental instabilities, or additional planets. $\delta_{\tau}$ is the Kronecker delta function, and $\tau$ represents a time interval between two measurements, $t-t'$. The white noise model uses log-normal priors, centred around the dispersion of the data and with a sigma similar to the dispersion of the data. This parametrization makes it difficult for the model to converge to very low white noise values, while not forbidding them, which helps prevent overfitting. 

The amplitudes $\alpha_{i}$ are related with the amplitude of the underlying function, not to any of the specific time series. These amplitudes are degenerate with the amplitudes of the component at each time series. We fixed $\alpha_{1}$ to 1, and let $\alpha_{2}$ free to account for scaling of the second oscillator, with a prior $\beta$ = $\mathcal{U}[0,1]$. 

We used a prior $\mathcal{U}[35,50]$ days for the rotation period, and a prior ln~$L$ $\mathcal{N}[4.5,0.8]$ days for the timescale of evolution \mbox{(90$^{+100}_{-50}$ days)}, following \citet{Giles2017}.

We used the S$_{MW}$  $\times$ 10 data, and photometry, as main dataset guiding the GP, which meant not including a gradient amplitude for them. The GP model, over all time series, is described as: 

\begin{equation} \label{eq_full_gp}
    \begin{split}
    \Delta ~S_{MW} \times 10 = &~A_{1}~k_{1} + \beta \cdot A_{1}~k_{2} ~, \\
    \Delta ~Flux = &~A_{2}~k_{1} + \beta \cdot A_{2}~k_{2} ~, \\
    \Delta ~FWHM = &~A_{3}~k_{1} + B_{3}~k'_{1} + \beta \cdot A_{3}~k_{2} + \beta \cdot B_{3}~k'_{2} ~, \\
    \Delta ~RV   = &~A_{4}~k_{1} + B_{4}~k'_{1} + \beta \cdot A_{4}~k_{2} + \beta \cdot B_{4}~k'_{2} ~, \\
    \end{split}
    \end{equation}

\noindent The scaling factor of the S$_{MW}$  $\times$ 10 ($A_{1}$) has a prior $\mathcal{U}[0,10]$, the scalling factor of the photometryc data ($A_{2}$) has a prior $\mathcal{U}[-20,20]$ ppt. The scaling factors of the FWHM and RV linear components ($A_{3}$ and $A_{4}$) have priors $\mathcal{U}[-50,50]$, while the scaling factors of the gradient components ($B_{3}$ and $B_{4}$) have priors $\mathcal{U}[-100,100]$.

\subsection{Correlation with the BERV}

There are several effects potentially affecting the spectra, such as telluric contamination, ghosts, or CCD stitching, anchored at the detector reference frame \citep{Dumusque2015, Coffinet2019, Cretignier2021}. Combined with the movement of the Earth around the Sun, this results in periodic shifts with 1-year periodicity. To account for potential contamination, we included a polynomial term against the BERV (see Fig. 3 of~\citealt{Dumusque2015}). This term is defined as shown in eq.~\ref{rv_berv_correl}, with different parameters for each instrument. We only found the BERV term to be significant for the HARPS data.

The correlation takes the form of: 
\begin{equation} \label{rv_berv_correl}
    \Delta RV = a \cdot BERV + b \cdot BERV^2
\end{equation}

\noindent with $a$ and $b$ being the parameters of the polynomial, with priors $\mathcal{N}[0,0.1]$. 

\subsection{HARPS focus drift}

HARPS had a known focus drift that manifested as a linear trend in FWHM data. The focus drift was corrected in 2015, when the fibers were upgraded. To account for this, we include a polynomial against time in the model of the FWHM of the HARPS-03. 

\begin{equation} \label{fwhm_time}
    \Delta FWHM = a \cdot (t - t_{0})
\end{equation}

\noindent with $a$ being the slope, with prior $\mathcal{N}[0,0.1]$. 

\subsection{Planetary model}

All previously stated components of the model define our null-hypothesis model, i.e. a model aimed to account for stellar activity and instrumental effect. On top of them, we include a planetary model where the number of planets will vary. The planetary signals are defined as circular for the detection tests, and later as full Keplerians to characterise their eccentricities. This choice avoids the potential pitfall of having a poorly sampled eccentric signal mimicking two circular signals and significantly cuts down on computing time. 

RV variations due to circular planetary orbits are defined as:
\begin{equation} \label{eq_circular}
    y(t)=-K \cdot \sin(2 \pi \cdot (t - t_{0})/P_{pl})
\end{equation} 

\noindent where $t_{0} = 7880 + P_{pl} \cdot (\phi_{pl} - 1)$. This parametrization ensures that our $t_{0}$ coincides to the inferior conjunction of the planets, and is within the baseline of observations. 

When conducting a blind search for planets, the orbital period is parametrised as angular frequency $\omega = 2\pi/P_{pl}$. We use a prior $\mathcal{U}[2\pi/1000$, $2\pi/2$]. When performing a guided search using the published solutions, we directly sample the periods and use $\mathcal{N}$ priors centred around the published solution. The phases $\phi_{pl}$ are parametrised as $\mathcal{U}[-0.25, 0.75]$. 

We parameterise the planetary amplitude $K$ as ln $K$ with a prior $\mathcal{U}[-5,2]$ m$\cdot$s$^{-1}$ (which keeps 50\% of the prior space below 22 cm$\cdot$s$^{-1}$). This parametrization expands the parameter space around amplitudes consistent with zero, reducing potential biases towards large posteriors in noise dominated data, as described in \citet{Rajpaul2024}. 

RV variations due to planetary elliptical orbits are defined as: 
\begin{equation} \label{eq_kepler}
    y(t)=K \left(\cos(\eta+\omega) + e \ \cos\omega\right)
\end{equation} 
  
\noindent where the true anomaly $\eta$ is related to the solution of the Kepler equation, which depends on the orbital period of the planet $P_{\rm orb}$ and the orbital phase $\phi$. This phase corresponds to the periastron time, which depends on the mid-point transit time $T_{0}$, the argument of periastron $\omega$, and the eccentricity of the orbit $e$.

We parametrise $e = (\sqrt{e} ~cos(\omega))^{2} + (\sqrt{e} ~sin(\omega))^{2}$ and $\omega = \arctan^{2}(\sqrt{e} ~sin(\omega),\sqrt{e} ~cos(\omega))$. We then sample $\sqrt{e} ~cos(\omega)$ and $\sqrt{e} ~sin(\omega)$ with priors $\mathcal{N}$[0, 0.3]. This parametrization favours low eccentricities, which are expected for short-period signals.

\clearpage
\section{Aliasing} \label{append_alias}

\begin{figure*}[!ht]
   \centering
  \includegraphics[width=18cm]{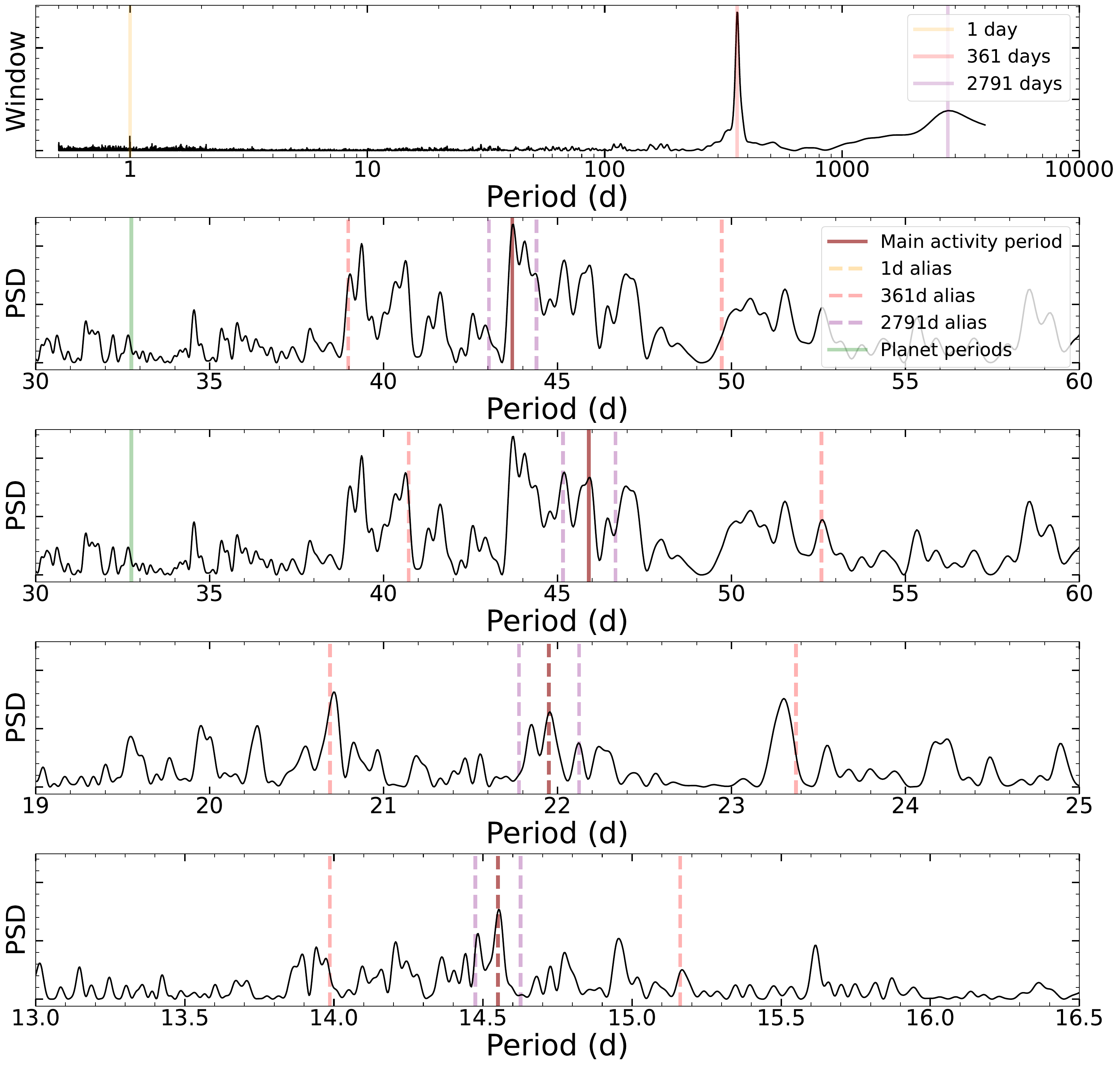}
  \caption{Aliasing of activity signals. The top panel shows the window function of the RV data, with the main peakis suspected of creating aliases hightlighted. The rest of the panels show the periodogram of the activity-only RVs, with the suspected peaks created by activity highlighted, along with their main aliases.}
  \label{window_alias}
\end{figure*}

\clearpage
\section{Adopted model -- Additional figures} \label{append_additional}

\begin{figure*}[!ht]
   \centering
  \includegraphics[width=18cm]{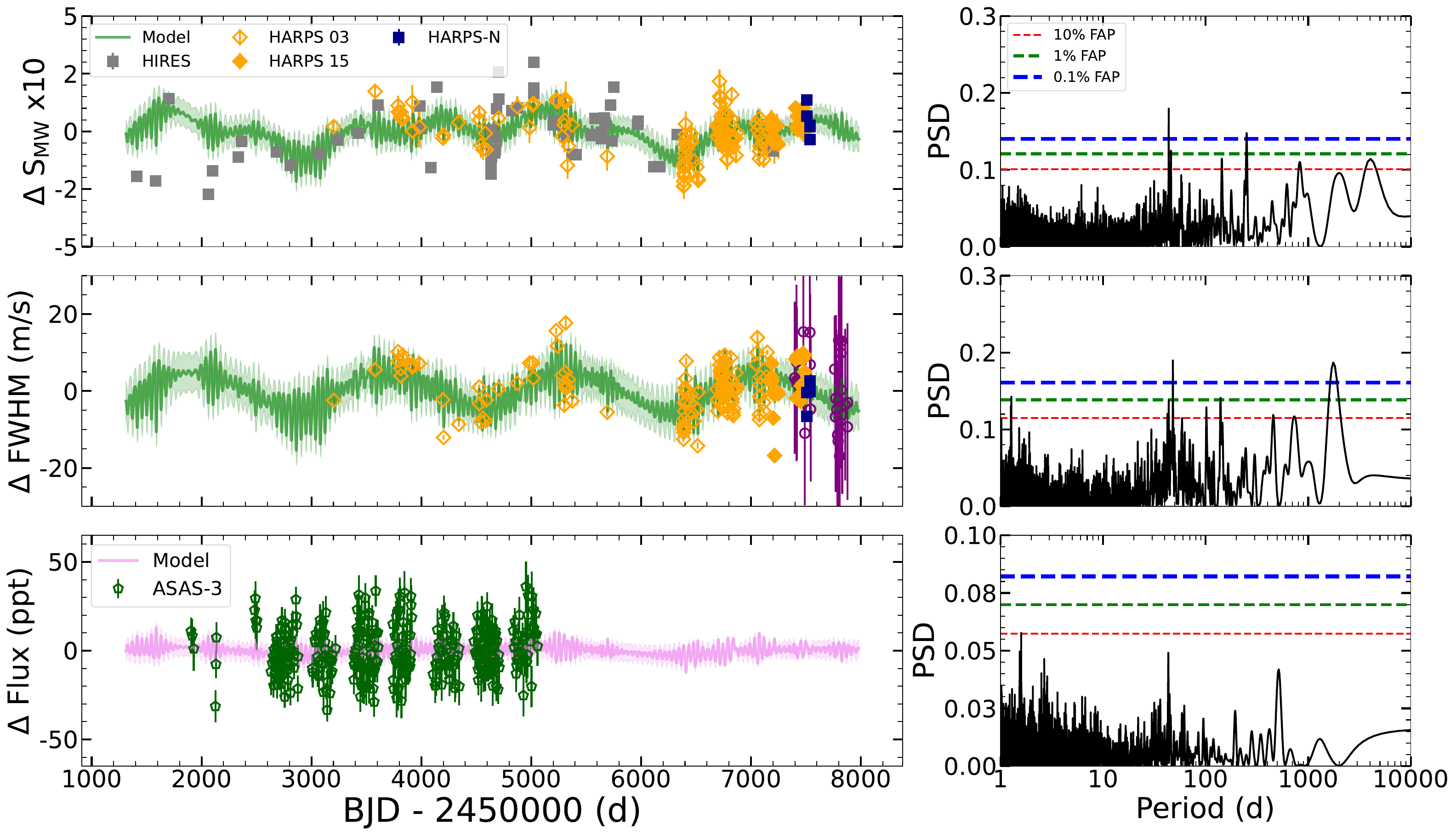}
  \caption{Adopted activity model. Activity proxies' data, with the best model fit, along with the periodogram of the data.}
  \label{adopted_act}
\end{figure*}

\begin{figure}[!ht]
   \centering
  \includegraphics[width=9cm]{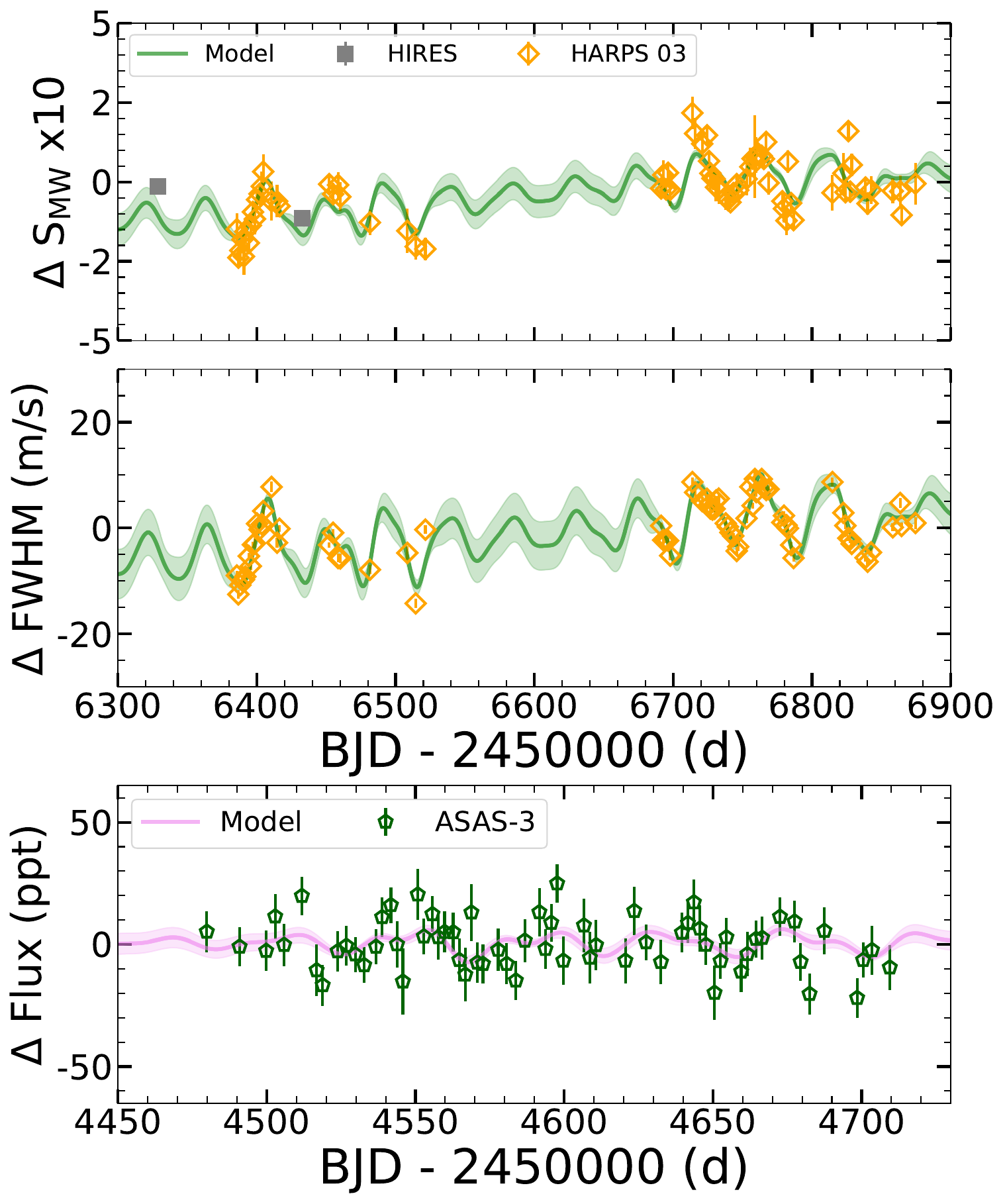}
  \caption{Zoom to selected observing campaigns. Activity proxies' data of with the best model fit.}
  \label{adopted_act_zoom}
\end{figure}

\clearpage
\section{Parameters of the model}\label{append_tables}



\renewcommand{\arraystretch}{1.36}
\begin{longtable}{llc} 
\caption{Priors and measured parameters of the adopted model.} \label{table_adopted} \\
\endfirsthead
\multicolumn{3}{c}{\tablename\ \thetable\ -- \textit{Continued from previous page}} \\
\hline
\endhead
\hline \multicolumn{3}{r}{\textit{Continued on next page}} \\
\endfoot
\hline
\endlastfoot
\hline
Parameter  & Priors & Posterior \\ \hline

\textbf{White noise} \\
ln $\sigma$ Phot$_{ASAS-3}$ [ppt] &  $\mathcal{N}$ (2.6 , 2.6) & 2.031$^{+0.093}_{-0.105}$ \\
\\
ln $\sigma$ S$_{MW~HARPS~03}$ $\times$ 10  &  $\mathcal{N}$ (0.15 , 0.15) & --0.483$^{+0.087}_{-0.085}$ \\
ln $\sigma$ S$_{MW~HARPS~15}$ $\times$ 10  &  $\mathcal{N}$ (0.15 , 0.15) & --0.15$^{+0.12}_{-0.13}$ \\
ln $\sigma$ S$_{MW~HARPS~N}$ $\times$ 10 &  $\mathcal{N}$ (0.15 , 0.15) & 0.10$^{+0.14}_{-0.14}$ \\
ln $\sigma$ S$_{MW~HIRES}$ $\times$ 10  &  $\mathcal{N}$ (0.15 , 0.15) & 0.00$^{+0.093}_{-0.095}$ \\

\\
ln $\sigma$ FWHM$_{HARPS~03}$  [m~s$^{-1}$] &  $\mathcal{N}$ (2 , 2) & 0.68$^{+0.13}_{-0.14}$ \\
ln $\sigma$ FWHM$_{HARPS~15}$  [m~s$^{-1}$] &  $\mathcal{N}$ (2 , 2) & 1.68$^{+0.19}_{-0.19}$ \\
ln $\sigma$ FWHM$_{HARPS~N}$ [m~s$^{-1}$]&  $\mathcal{N}$ (2 , 2) & 1.30$^{+0.52}_{-0.45}$ \\
ln $\sigma$ FWHM$_{CARMENES}$ [m~s$^{-1}$]&  $\mathcal{N}$ (2 , 2) & 0.4$^{+1.0}_{-1.4}$ \\

\\ 
ln $\sigma$ RV$_{HARPS~03}$ [m~s$^{-1}$]  &  $\mathcal{N}$ (1 , 1) & --0.37$^{+0.25}_{-0.34}$ \\
ln $\sigma$ RV$_{HARPS~15}$ [m~s$^{-1}$]  &  $\mathcal{N}$ (1 , 1) & 0.35$^{+0.24}_{-0.27}$ \\
ln $\sigma$ RV$_{HARPS~N}$ [m~s$^{-1}$]  &  $\mathcal{N}$ (1 , 1)& 0.45$^{+0.44}_{-0.44}$ \\
ln $\sigma$ RV$_{CARMENES}$ [m~s$^{-1}$]  &  $\mathcal{N}$ (1 , 1) & --0.28$^{+0.46}_{-0.60}$ \\
ln $\sigma$ RV$_{HIRES}$ [m~s$^{-1}$]  &  $\mathcal{N}$ (1 , 1) & 0.69$^{+0.19}_{-0.22}$ \\
\\
\textbf{Zero points} \\
V0 Phot$_{ASAS-3}$ [ppt] &  $\mathcal{N}$ (0 , 10) &   1.43$^{+0.73}_{-0.71}$ \\
\\
V0 S$_{MW~HARPS~03}$ $\times$ 10  &  $\mathcal{N}$ (0 , 10) & 0.17$^{+0.11}_{-0.11}$\\
V0 S$_{MW~HARPS~15}$ $\times$ 10 &  $\mathcal{N}$ (0 , 10) & --0.94$^{+0.26}_{-0.26}$ \\
V0 S$_{MW~HARPS~N}$ $\times$ 10 &  $\mathcal{N}$ (0 , 10) & 0.16$^{+0.57}_{-0.57}$\\
V0 S$_{MW~HIRES}$ $\times$ 10 &  $\mathcal{N}$ (0 , 10) & 1.35$^{+0.15}_{-0.15}$\\

\\
V0 FWHM$_{HARPS~03}$ [m~s$^{-1}$] &  $\mathcal{N}$ (0 , 10) & --6.6$^{+1.3}_{-1.4}$ \\
V0 FWHM$_{HARPS~15}$ [m~s$^{-1}$] &  $\mathcal{N}$ (0 , 10) & --2.5$^{+1.7}_{-1.7}$\\
V0 FWHM$_{HARPS~N}$ [m~s$^{-1}$]&  $\mathcal{N}$ (0 , 10) & 1.5$^{+3.2}_{-3.4}$\\
V0 FWHM$_{CARMENES}$ [m~s$^{-1}$]&  $\mathcal{N}$ (0 , 10) & 2.2$^{+3.8}_{-3.6}$ \\
\\
V0 RV$_{HARPS~03}$ [m~s$^{-1}$] &  $\mathcal{N}$ (0 , 10) & --1.2$^{+0.25}_{-0.25}$\\
V0 RV$_{HARPS~15}$ [m~s$^{-1}$] &  $\mathcal{N}$ (0 , 10) & 0.47$^{+0.57}_{-0.57}$\\
V0 RV$_{HARPS~N}$ [m~s$^{-1}$] &  $\mathcal{N}$ (0 , 10) & 1.7$^{+1.2}_{-1.2}$\\
V0 RV$_{CARMENES}$ [m~s$^{-1}$] &  $\mathcal{N}$ (0 , 10) & 0.13$^{+0.51}_{-0.52}$\\
V0 RV$_{HIRES}$ [m~s$^{-1}$] &  $\mathcal{N}$ (0 , 10) & 1.02$^{+0.42}_{-0.44}$\\

\\
\textbf{Polynomial parameters} \\
a FWHM$_{HARPS~03}$ [m~s$^{-1}$ d$^{-1}$] &  $\mathcal{N}$ (0 , 0.1) & --0.00184$^{+0.00051}_{-0.00052}$ \\ 
\\
a BERV$_{HARPS}$ [m~s$^{-1}$ / km~s$^{-1}$] &  $\mathcal{N}$ (0 , 0.1) & 0.0036$^{+0.0073}_{-0.0073}$ \\
b BERV$_{HARPS}$ [m~s$^{-1}$ / km$^{2}$~s$^{-2}$] &  $\mathcal{N}$ (0 , 0.1) & 0.00293$^{+0.00045}_{-0.00045}$ \\
\\
\textbf{Cycle parameters} \\
Ln Period [d] &  $\mathcal{U}$ (7 , 8.7) & 8.128$^{+0.032}_{-0.018}$ \\ 
Phase$_{P}$    & $\mathcal{N}$ (-0.5 , 1) & 0.315$^{+0.054}_{-0.045}$\\
Phase$_{P/2}$    & $\mathcal{N}$ (-0.5 , 1) & --0.229$^{+0.075}_{-0.052}$\\
Phase$_{P/4}$    & $\mathcal{N}$ (-0.5 , 1) & --0.060$^{+0.054}_{-0.047}$\\
\\
K Phot$_{P}$ [ppt] &  $\mathcal{N}$ (0 , 10) &  1.62$^{+0.95}_{-0.91}$\\
K Phot$_{P/2}$ [ppt] &  $\mathcal{N}$ (0 , 10) & 0.95$^{+0.97}_{-0.97}$ \\
K Phot$_{P/4}$ [ppt] &  $\mathcal{N}$ (0 , 10) & 0.17 $^{+0.92}_{-0.92}$ \\
\\
K S$_{MW}$ $\times$ 10 $_{P}$ &  $\mathcal{U}$ (0 , 2) &  0.46$^{+0.12}_{-0.12}$\\
K S$_{MW}$ $\times$ 10 $_{P/2}$&  $\mathcal{N}$ (0 , 10) & 0.36$^{+0.11}_{-0.11}$  \\
K S$_{MW}$ $\times$ 10 $_{P/4}$ &  $\mathcal{N}$ (0 , 10) & 0.394$^{+0.092}_{-0.092}$ \\
\\
K FWHM$_{P}$ [m~s$^{-1}$] &  $\mathcal{N}$ (0 , 10) &  0.62$^{+0.87}_{-0.89}$\\
K FWHM$_{P/2}$ [m~s$^{-1}$] &  $\mathcal{N}$ (0 , 10) & 4.71$^{+0.69}_{-0.67}$  \\
K FWHM$_{P/4}$ [m~s$^{-1}$] &  $\mathcal{N}$ (0 , 10) & 0.58$^{+0.71}_{-0.71}$ \\

\\
K RV$_{P}$ [m~s$^{-1}$] &  $\mathcal{N}$ (0 , 10) & --0.06$^{+0.29}_{-0.28}$ \\
K RV$_{P/2}$ [m~s$^{-1}$] &  $\mathcal{N}$ (0 , 10) & --0.10$^{+0.25}_{-0.25}$ \\
K RV$_{P/4}$ [m~s$^{-1}$] &  $\mathcal{N}$ (0 , 10) & 1.36$^{+0.21}_{-0.22}$ \\

\\
\textbf{GP Parameters} \\
Rotation period [d] &  $\mathcal{U}$ (35 , 50) & 43.63$^{+0.85}_{-0.72}$ \\ 
ln Timescale [d] &  $\mathcal{N}$ (4.5, 0.8) & 4.38$^{+0.40}_{-0.38}$ \\ 
$\beta$ &  $\mathcal{U}$ (0, 1) & 0.430$^{+0.097}_{-0.085}$ \\
A Phot [ppt] &  $\mathcal{U}$ (-20 , 20) & -4.34$^{+0.95}_{-1.00}$ \\
A S$_{MW}$ $\times$ 10 $_{P}$ &  $\mathcal{U}$ (0 , 10) &  0.513$^{+0.091}_{-0.080}$\\
A FWHM[m~s$^{-1}$] &  $\mathcal{U}$ (-50 , 50) & 4.55$^{+0.63}_{-0.53}$ \\
B FWHM [m] &  $\mathcal{U}$ (-100 , 100) & --5.3$^{+2.9}_{-3.1}$ \\
A RV [m~s$^{-1}$] &  $\mathcal{U}$ (-50 , 50) & 1.25$^{+0.23}_{-0.21}$ \\
B RV [m] &  $\mathcal{U}$ (-100 , 100) &  10.0$^{+1.7}_{-1.5}$\\

\\
\textbf{Planets} \\
ln K$_{b}$ [m~s$^{-1}$] &  $\mathcal{U}$ (-5 , 2) & 1.109$^{+0.047}_{-0.048}$\\
Period$_{b}$ [d] &  $\mathcal{U}$ (8.95,10.45) & 8.70874$^{+0.00056}_{-0.00055}$ \\
Phase$_{b}$  &  $\mathcal{U}$ (-0.25 , 0.75) & 0.208 $^{+0.014}_{-0.014}$ \\
\\
ln K$_{c}$ [m~s$^{-1}$] &  $\mathcal{U}$ (-5 , 2) & 0.59$^{+0.11}_{-0.12}$ \\
Period$_{c}$ [d] &  $\mathcal{U}$ (26.2,39.2) & 32.761$^{+0.015}_{-0.016}$ \\
Phase$_{c}$  &  $\mathcal{U}$ (-0.25 , 0.75) & 0.614 $^{+0.028}_{-0.029}$ \\

\hline
\end{longtable}

\clearpage
\section{TESS photometry}\label{tess_phot}

\begin{figure}[h]
  \centering
     \includegraphics[width=18cm]{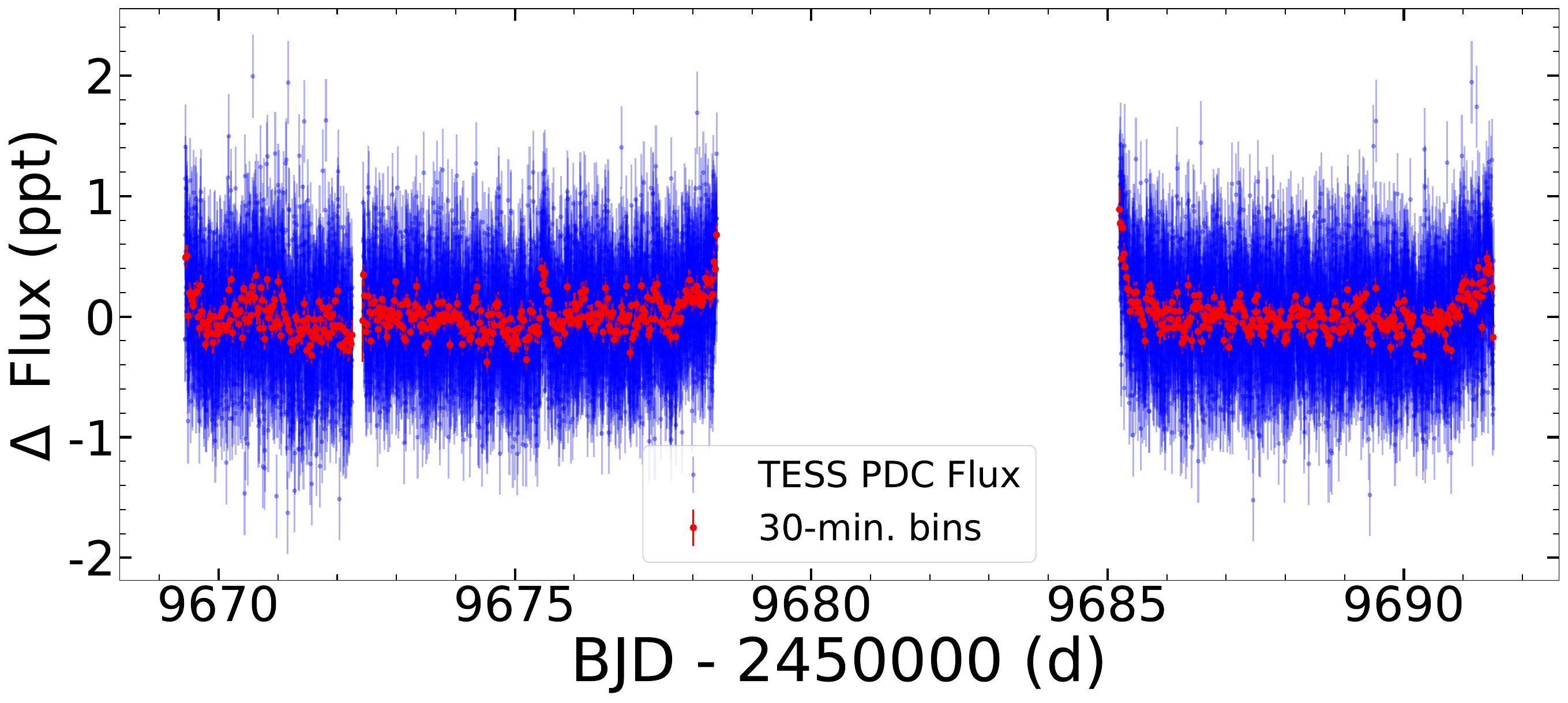}
	\caption{\textit{TESS} light curve. Blue symbols show the 2-minute cadence data. The red symbols represent 30-minute bins.}

	\label{tess_data}
\end{figure}

GJ~536 (TIC 119147875) was observed by the Transiting Exoplanet Survey Satellite (\textit{TESS}) \citep{Ricker2015} with the 2\,min cadence in sector 50. We analysed the pre-search data conditioned simple aperture photometry (PDCSAP) fluxes \citep{Smith2012, Stumpe2012}. The PDCSAP light curve shows an RMS of 400 ppm (155 ppm in 30-minute bins). Figure~\ref{tess_data} shows the \textit{TESS} PDCSAP light curve. No aparent features consistent with transits appear in the data. We applied the box least squares (BLS) periodogram \citep{Kovacs2002, Hartman2016} to the \textit{TESS} time series data to search for transit features. No transits have been found in the BLS periodogram. 

\end{appendix}
\label{lastpage}

\end{document}